\documentclass[%
 aip,
 amsmath,amssymb,
reprint,%
]{revtex4-2}
\usepackage{amsmath,amssymb}
\usepackage{graphicx}
\usepackage{dcolumn}
\usepackage{bm}

\usepackage[utf8]{inputenc}
\usepackage[T1]{fontenc}
\usepackage{mathptmx}
\usepackage{etoolbox}
\usepackage{xcolor}
\usepackage{graphicx}
\usepackage{subcaption}
\usepackage{caption}
\usepackage{xcolor,soul} 
\sethlcolor{yellow} 

\makeatletter
\def\@email#1#2{%
 \endgroup
 \patchcmd{\titleblock@produce}
  {\frontmatter@RRAPformat}
  {\frontmatter@RRAPformat{\produce@RRAP{*#1\href{mailto:#2}{#2}}}\frontmatter@RRAPformat}
  {}{}
}%
\makeatother
\begin{document}

\preprint{AIP/123-QED}

\title{Multiscale Convolutional Neural Networks for Subgrid-scale Modeling in Large-Eddy Simulation}
\author{Bahrul Jalaali$^*$}
    \affiliation{ 
Department of Mechanical Engineering, Osaka University, 2-1 Yamadaoka, Suita, Osaka, 565-0871, Japan
}%
\author{Kie Okabayashi}%
    \affiliation{ 
Department of Mechanical Engineering, Osaka University, 2-1 Yamadaoka, Suita, Osaka, 565-0871, Japan
}%
\email{jalaali@fluid.mech.eng.osaka-u.ac.jp} 

\date{\today}

\begin{abstract}
This study proposes a multiscale convolutional neural network subgrid-scale (MSC-SGS) model for large-eddy simulation (LES). This model incorporates multiscale representations obtained via filtering to capture turbulent vortices interactions and physical processes at different scales. Subsequently, it progressively encodes information from the largest to the smallest scale, thereby mimicking an energy-cascade process. A turbulent channel flow with $\text{Re}_{\tau} = 180$  is adopted as the training and testing dataset, whereas the rate-of-strain tensor is used as the input variable to adhere the rotational invariance. \textit{A priori} test results show that the MSC-SGS model predicts the physical quantities of residual stress, SGS dissipation, backscatter, and SGS transport more accurately than two other convolutional neural network (CNN)-based monoscale and U-Net, while maintaining high temporal correlation. Based on \textit{a posteriori} test, the MSC-SGS model outperforms the two other CNN models and the conventional Smagorinsky model in predicting turbulence statistics while ensuring numerical stability without resorting to ad-hoc treatments such as clipping excessive backscatters. The LES results based on the MSC-SGS model closely align with DNS data in terms of turbulence statistics, energy spectra, and the quantitative reproduction of instantaneous-flow structures. These findings suggest that incorporating multiscale representations effectively advances the development of SGS models for LES.
\end{abstract}

\maketitle

\section{Introduction}
In recent years, researchers have attempted to construct data-driven turbulence models that replace mathematical turbulence models with deep neural networks (DNN) trained on a database of high-fidelity numerical simulations. Unlike conventional deductive mathematical models, data-driven models can inductively extract subfilter-scale fluctuations and create phenomenon-based models that do not contain artificial approximations or assumptions when trained in appropriate settings. One such application of the data-driven approach is the data-driven subgrid-scale (SGS) model within large-eddy simulation (LES) frameworks. The data-driven SGS model is designed to be constructed between a grid-scale (GS) resolved flow and SGS unclosed terms for predicting residual stress or SGS stress ($\tau_{ij}$) using DNN trained on space-filtered direct numerical simulation (DNS). Furthermore, Duraisamy et al.\cite{duraisamy_turbulence_2019} and Vinuesa \& Brunton\cite{vinuesa_emerging_2022} provided extensive reviews of data-driven turbulence models, which highlighted their potential as a feasible solution to improve the representation of SGS terms.

Efforts have been devoted to developing data-driven SGS models using DNN, in particular using fully connected multilayer perceptron (MLP). This approach involves a nonlinear mapping between the resolved flow fields from a filtered DNS (fDNS) and an unresolved $\tau_{ij}$ stress via a series of matrix multiplications and nonlinear activation functions. For instance, Gamahara \& Hattori\cite{gamahara_searching_2017} predicted $\tau_{ij}$ in a channel flow using the velocity gradient tensor $\frac{\partial \bar{u}_i}{\partial x_j}$ and the distance from the channel walls as the input for their data-driven SGS model. Their \textit{a priori} test results showed a high correlation coefficient between the actual $\tau_{ij}$ calculated from fDNS data and the predicted $\tau_{ij}$. However, the model did not outperform the conventional Smagorinsky model (SMAG) in actual LES computations (\textit{a posteriori} test). Liu et al.\cite{liu_investigation_2022} employed a data-driven SGS model using velocity field as the input to predict $\tau_{ij}$. Their \textit{a priori} test results aligned well with fDNS data, although in \textit{a posteriori} tests their model yielded slightly overestimated mean flow values. For the case of isotropic turbulence, Zhou et al.\cite{zhou_subgrid-scale_2019} introduced a data-driven SGS model that uses $\frac{\partial \bar{u}_i}{\partial x_j}$ and the filter width as inputs. Similar to the findings of Gamahara \& Hattori\cite{gamahara_searching_2017} and Liu et al.\cite{liu_investigation_2022}, this model aligned well with fDNS data in \textit{a priori} test. Nevertheless, this model did not demonstrate any clear advantage over conventional SGS models in \textit{a posteriori} test, and the reasons remain uncertain. Data-driven SGS models\cite{gamahara_searching_2017,liu_investigation_2022,zhou_subgrid-scale_2019}  utilize MLP in a pointwise manner, in which single-point information is solely considered in the input and output prediction. Subsequent studies have investigated the use of multipoint inputs instead of the single inputs to enhance predictive accuracy. This approach allows a grid of stencil points to be assigned to the input variable in order to capture additional spatial information. Xie et al.\cite{xie_modeling_2020} employed a data-driven SGS model for isotropic turbulence with multipoint $\frac{\partial \bar{u}_i}{\partial x_j}$ at different spatial location. The results yielded accurate predictions of SGS terms in \textit{a priori} test, while showing improvement over the conventional LES model in \textit{a posteriori} test. For a similar isotropic turbulence case, Maulik et al.\cite{maulik_sub-grid_2019} used an MLP with the vorticity ($\omega$) and the stream function ($\psi$) as the multipoint inputs. The results demonstrated the advantage of the data-driven model over the conventional Smagorinsky and Leith models. Additionally, Park \& Choi\cite{park_toward_2021} tested both single- and multipoint resolved rate-of-strain tensors and $\frac{\partial \bar{u}_i}{\partial x_j}$ as inputs for a data-driven SGS model in a channel flow. They highlighted that introducing multipoint inputs provided higher accuracy in predicting $\tau_{ij}$. 

Despite such potentially favorable features, the MLP data-driven approach with multiple inputs results in heavy computational burden, which is less effective for scenarios involving big data such as turbulent flow fields\cite{liu_investigation_2022}. An effective method to extract spatial information is using a convolutional neural network (CNN), which captures spatial information through convolutional kernels. Consequently, the memory allocation of DNN parameters and the computational time are reduced significantly, thus rendering it suitable for flow-field data. Owing to these advantages, Liu et al.\cite{liu_investigation_2022} compared the MLP and CNN approaches on a data-driven SGS model in a turbulent channel. They constructed a data-driven SGS model to predict $\tau_{ij}$ using velocity components as the input and applied a two-dimensional (2D) convolutional kernel to the local 2D velocity field on an $x-z$ plane. Their findings revealed that the CNN-based data-driven model achieved better results than the MLP-based model in \textit{a posteriori} test, where more accurate results with improved numerical stability were yielded. In contrast, MLP-based data-driven models typically required additional treatment, such as clipping (as demonstrated by Zhou et al.\cite{zhou_subgrid-scale_2019}, Maulik et al.\cite{maulik_sub-grid_2019} and Park \& Choi\cite{park_toward_2021}), to maintain stability in \textit{a posteriori} tests. Similarly, Guan et al.\cite{guan_stable_2022} obtained stable results using a 2D-CNN data-driven SGS model in homogeneous isotropic turbulence using $\omega$ and $\psi$ as input variables. They reported that, without any treatment in \textit{a posteriori} tests, the data-driven model offered accurate predictions that aligned closely with fDNS data. Moreover, by considering the three-dimensional (3D) characteristics of turbulent flows, Saura \& Gomez\cite{saura_predicting_2023} argued that a 3D-CNN was more favorable for capturing flow-field information than a 2D-CNN. They employed a 3D-CNN based on the U-Net architecture to predict $\tau_{ij}$ using velocity as the input, which showed a high correlation with fDNS data in \textit{a priori} test, whereas \textit{a posteriori} results were not evaluated in their study. Beck et al.\cite{beck_deep_2019} constructed a 3D-CNN with a residual network structure to predict unresolved terms in LES. In contrast to the results of Saura \& Gomez\cite{saura_predicting_2023}, their \textit{a priori} result depicted a decent correlation between fDNS data. However, based on \textit{a posteriori} result, the data-driven model outperformed the conventional Smagorinsky method in terms of turbulent statistics. This finding underscores the superiority of 3D-CNN models in accurately predicting unresolved terms during \textit{a posteriori} tests. 

While both MLP- and CNN-based approaches have shown potential, capturing spatial features in the multiscale dynamics of turbulent fields remains a challenge in order to accurately predict $\tau_{ij}$. To ensure a successful data-driven SGS model, capturing the multiscale characteristics of turbulent flow dynamics is essential, as highlighted by Xie et al.\cite{xie_modeling_2020} The multiscale interaction of vortices, in which energy and momentum are transferred across a wide range of scales, is expected to be accurately captured for reliable predictions in SGS models because it affects the overall flow dynamics. CNN models, owing to their inherent ability to capture spatial hierarchies through convolutional layers, are particularly suitable for this task. Moreover, data-driven SGS models based on MLP typically exhibit unstable results in \textit{a posteriori} tests, as mentioned in a review by Cinnella\cite{cinnella_data-driven_2024}. Similarly, Park \& Choi\cite{park_toward_2021} reported that a high correlation between predicted $\tau_{ij}$ and fDNS data in \textit{a priori} tests did not guarantee the \textit{a posteriori} results. This inconsistency suggests the limited generalizability for unseen data in \textit{a posteriori} tests, which has been observed in conventional SGS models, as reported by Clark et al.\cite{clark_evaluation_1979} and Park et al.\cite{park_toward_2005} Gamahara \& Hattori\cite{gamahara_searching_2017} employed an MLP-based SGS model and observed that the data-driven SGS model behaved similarly to a structural SGS model in an LES. Structural SGS models show good correlations in \textit{a priori} tests but tend to exhibit insufficient dissipation and destabilize the numerical calculations in \textit{a posteriori} tests\cite{kajishima_computational_2017}. Hence, Kang et al.\cite{kang_neural-network-based_2023} introduced a mixed model comprising an MLP-based SGS that incorporates the rate-of-strain tensor and resolved stress $L_{ij} = \widehat{\overline{u}_i\overline{u}_j} - \widehat{\overline{u}_i} \widehat{\overline{u}_j}$ as inputs to enhance dissipation on a local grid, where $\hat{(\cdot)}$ denotes the test filter. Providing an additional input variable yielded \textit{a posteriori} results that closely aligned with fDNS data without ad-hoc treatment, although reliance on resolved stress data remains a disadvantage owing to their low availability. By contrast, some studies\cite{liu_investigation_2022, guan_stable_2022, beck_deep_2019} demonstrated that CNN-based data-driven SGS models can perform well in \textit{a posteriori} tests without any additional treatment or additional input, thus demonstrating their ability to capture dissipation effectively. By leveraging the ability of CNN to capture multiscale spatial features, the inaccuracies observed in \textit{a posteriori} results can be overcome.

To advance the development of CNN-based SGS models, this study focuses on the multiscale nature of turbulence vortices and proposes a novel approach for the multiscale-CNN SGS model (MSC-SGS model). Previous multiscale CNN approaches incorporated different scales of inputs, either by using a distinct CNN kernel (Yuan et al.\cite{yuan_multiscale_2018}) or by applying a downsampling process (Illaramendi et al.\cite{ajuria_illarramendi_performance_2022}); however, they generally required a single-input representation. In the MSC-SGS model, multiscale representations are applied to the data-driven SGS model via filtering to ensure that the model captures intricate details across scales. Inspired by the multifidelity technique, the MSC-SGS model progressively captures features from the largest to the smallest scales, thereby mimicking the energy-cascade process in turbulent flows. Moreover, this study aims to examine whether the proposed model can extract features of complex turbulent flow fields with turbulent vortical structures in various scales more effectively than the conventional CNN-based model, as well as to assess its effect on the prediction of $\tau_{ij}$. The remainder of this paper is organized as follows: The numerical method and methodology pertaining to the DNN SGS model are explained in Section II. The details of \textit{a priori} and \textit{a posteriori} tests on an actual LES are discussed in Section III and IV, respectively. The concluding remarks are presented in Section V.

\section{Methodology}
    \subsection{\label{sec:level2}Problem setting}
    In an LES, the turbulent channel flow between two parallel flat plates driven by a constant pressure gradient is simulated by solving the spatially filtered forms of the continuity and Navier$\--$Stokes equations as follows:

    \begin{equation}
        \frac{\partial \bar{u}_i}{\partial x_i} = 0
        \label{mass}
    \end{equation}
    
    \begin{equation}
        \frac{\partial \bar{u}_i}{\partial t} + \frac{\partial (\bar{u}_i \bar{u}_j)}{\partial x_j} = - \frac{\partial \bar{p}}{\partial x_i} + \frac{1}{\text{Re}} \frac{\partial}{\partial x_j} \left(-\tau_{ij} + 2\bar{D}_{ij}\right)
        \label{momentum}
    \end{equation}
    
    where the overbar denotes the filtering operation, and $\bar{D}_{ij}$ denotes the grid-scale (GS) rate-of-strain tensor:
    
    \begin{equation}
        \bar{D}_{ij} = \frac{1}{2} \left( \frac{\partial \bar{u}_i}{\partial x_j} + \frac{\partial \bar{u}_j}{\partial x_i} \right)
        \label{dij}
    \end{equation}
    
\begin{figure}
    \includegraphics[width=1\linewidth]{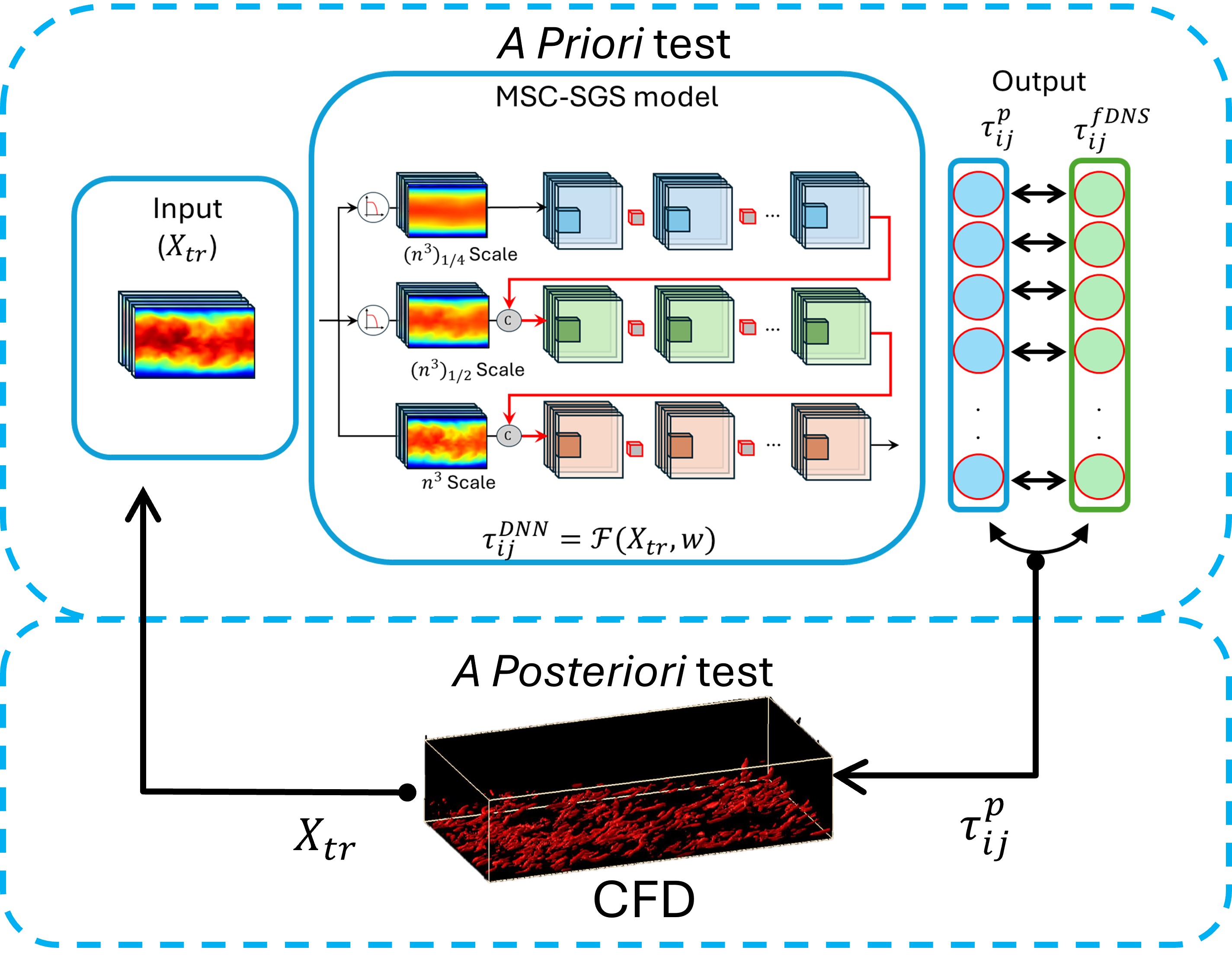}
    \caption{\label{Fig1} Schematic diagram of data-driven SGS framework.}
    \end{figure}

    The residual SGS stress, $\tau_{ij} = \overline{u_i u_j} - \bar{u}_i \bar{u}_j$, is unclosed and should be modeled. In this study, the data-driven SGS model ($\tau_{ij}^{DNN}$) is introduced to model the unclosed $\tau_{ij}$, thus replacing the conventional SGS approach. Fig. \ref{Fig1} shows a schematic representation of the proposed methodology for constructing a $\tau_{ij}^{DNN}$ model. 
    
    In \textit{a priori} test, the $\tau_{ij}^{DNN}$ model is trained in a supervised manner using a dataset of flow-field variables from DNS that comprises a set of input data $X_{tr}$ and label data $S_{tr}$. This step aims to establish a functional relation between the input and output variables, thus yielding a nonlinear regression $\tau_{ij}^{DNN} = \mathcal{F}(X_{tr}, w)$,    
    where $w$ indicates the weight of the neural network. The training process can be regarded as an optimization problem to determine the weight such that $w = \arg \min_w \left( L(S_{tr}, \tau_{ij}^{p}) \right)$, where $L$ denotes the loss function and $\tau_{ij}^{p}$ is the predicted $\tau_{ij}$. Subsequently, the trained model of $\tau_{ij}^{DNN}$ is implemented in \textit{a posteriori} test to assess its performance within the actual LES computation.
    
    \subsection{\label{sec:level2}Preparation of training dataset}
\begin{figure}
    \includegraphics[width=1\linewidth]{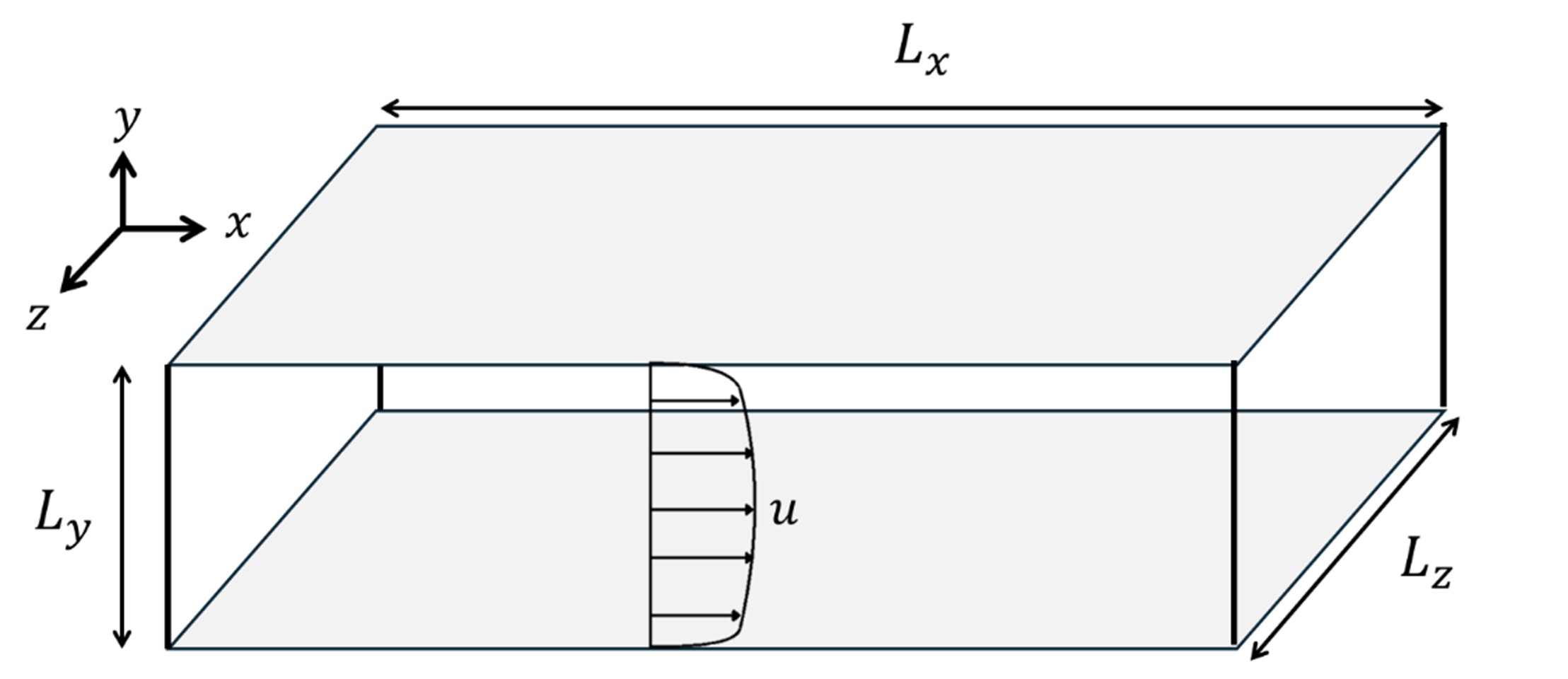}
    \caption{\label{Fig2} Computational domain.}
    \end{figure}

\begin{figure*} 
    \centering
        \begin{subfigure}{0.45\textwidth} 
            \centering
            \includegraphics[width=\linewidth]{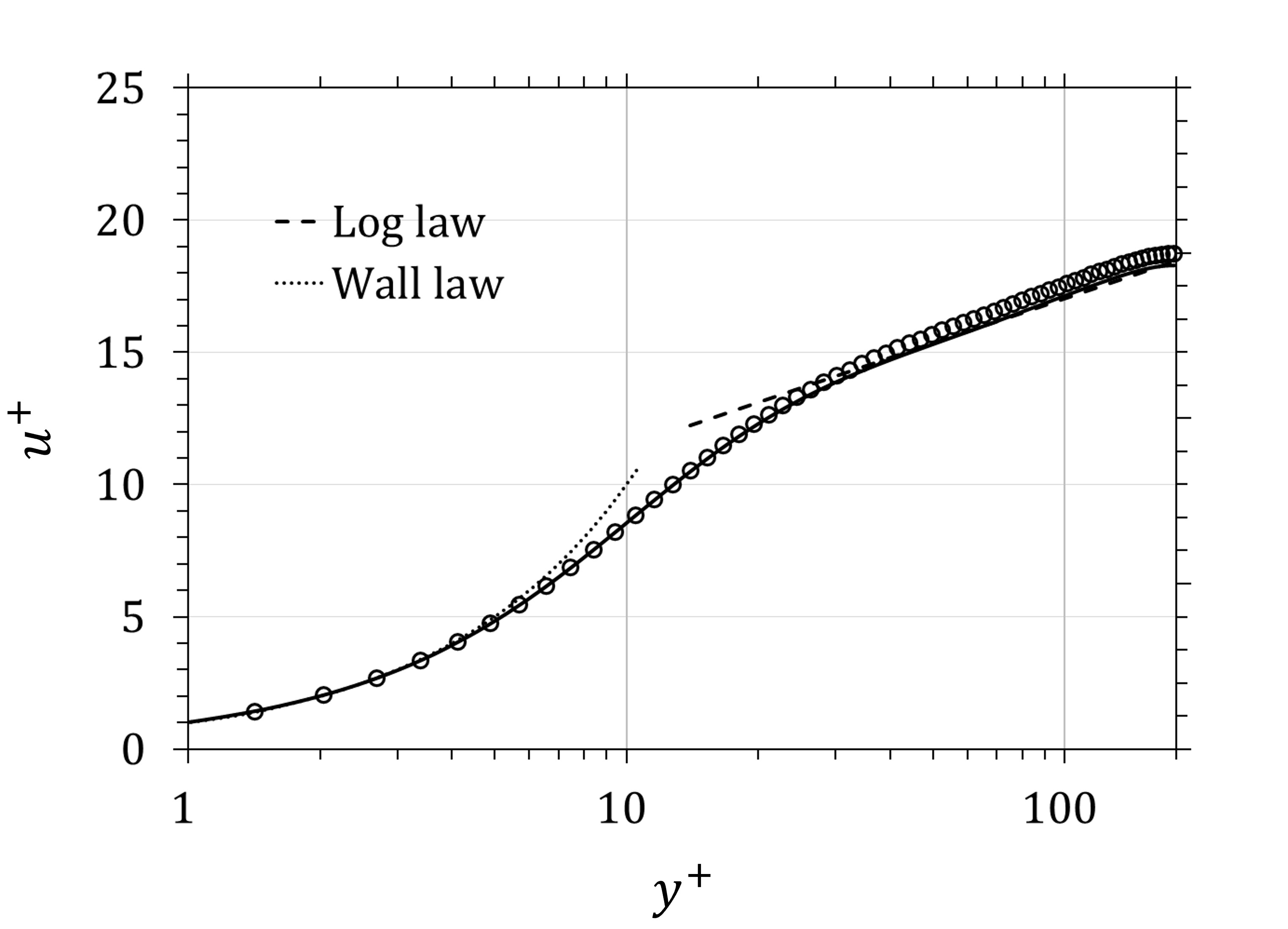}
            \caption{}
            \label{Fig3a}
        \end{subfigure}
        \hfill
        \begin{subfigure}{0.45\textwidth}
            \centering
            \includegraphics[width=\linewidth]{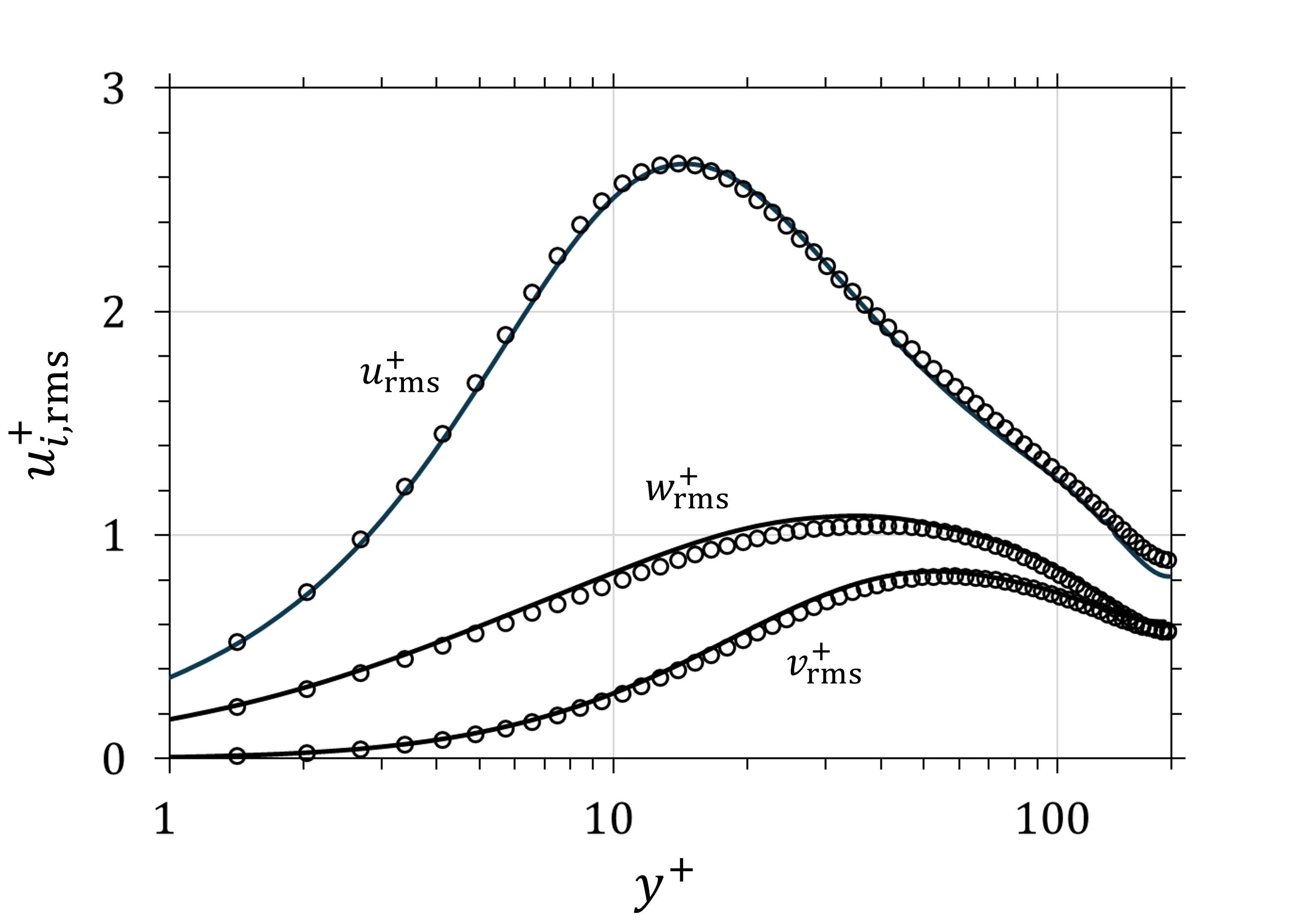}
            \caption{}
            \label{Fig3b}
        \end{subfigure}
        \caption{Wall-normal distribution of turbulence statistics, where ``o'' shows the present results and the solid line corresponds to the results of Kim et al.~\cite{kim_turbulence_1987} (a) mean velocity, and (b) Root-mean-square velocity.}
        \label{Fig3}
    \end{figure*}

    The dataset used to develop the data-driven SGS model is obtained using the filtered DNS (fDNS) data of a turbulent channel flow with a constant pressure gradient in the $x$ direction, which is essentially the same setting as the \textit{a posteriori} LES computation. The governing equations are the continuity equation and Navier$-$Stokes equations for incompressible flow, and the computational domain is depicted in Fig. \ref{Fig2}. The domain dimensions and grid points are  $(L_x \times L_y \times L_z) = (4\pi\delta \times 2\delta \times 2\pi\delta)$ and $(N_x \times N_y \times N_z) = (192 \times 128 \times 160)$, respectively, where $\delta$ is the channel half-width. The grid spacing in each direction is similar to that used by Kim et al.\cite{kim_turbulence_1987} The collocation grid is employed with a non-uniform grid in the wall-normal ($y$) direction, whereby the no-slip boundary condition is applied to the walls. Periodicity is assumed in the streamwise and spanwise directions.The unsteady simulation method is based on the fractional step method, with convective and viscous terms discretized using a second-order central finite difference scheme. The Adams$-$Bashforth method is adopted for time marching. The friction Reynolds number $\text{Re}_{\tau} = \frac{u_{\tau} \delta}{\nu}$ is set as 180, where $u_{\tau}$ and $\nu$ denote the frictional mean velocity and kinematic viscosity, respectively. The validity of the DNS result is depicted in Fig. \ref{Fig3}, which shows the wall-normal distributions of the mean streamwise velocity ($u^+$) and the root-mean-square (rms) values of the fluctuating velocity components in streamwise ($u_{\mathrm{rms}}^+$), wall-normal ($v_{\mathrm{rms}}^+$), and spanwise ($w_{\mathrm{rms}}^+$) directions. These results agree well with those reported by Kim et al.~\cite{kim_turbulence_1987}

    Next, the DNS flow field data are filtered to separate the GS and SGS variables, which denotes as filtered DNS (fDNS). The GS variables are used as input, whereas the SGS stress, $\tau_{ij}^{fDNS} = \overline{u_i u_j} - \overline{u}_i \overline{u}_j$, is calculated to provide the corresponding label data ($S_{tr}$). A box filter is applied as described in Eq.~\eqref{filter}.
    
    \begin{equation}
    G(x) = \begin{cases} 
        1 / \bar{\Delta} & (|x| < \bar{\Delta}/2) \\
        0 & (|x| > \bar{\Delta}/2)
        \label{filter}
    \end{cases} 
    \end{equation}
    
    In this study, the filter sizes in each direction are $\bar{\Delta}_x = 6\Delta_x^{DNS}$, $\bar{\Delta}_y = 2\Delta_y^{DNS}$ and $\bar{\Delta}_z = 5\Delta_z^{DNS}$, respectively ($\Delta$: grid size). Filtering is conducted to ensure the compatibility of the $\tau_{ij}^{DNN}$ model to be trained in a coarser grid, which is similar to the resolution of an actual LES. The dataset comprises 140 instantaneous fields, which correspond to approximately 1,820 non-dimensional time units ($tU_{bulk}/\delta$), 25,180 viscous time scales ($t u_{\tau}^2 / \nu$), and 291 flow throughs ($tU_{bulk} /L_z$), whereby $U_{bulk}$ and $u_{\tau}$ denote the bulk and friction velocities, respectively. Additionally, 80\% and 20\% of the dataset are used for training and testing, respectively. The components of $\tau_{ij}^{fDNS}$, averaged in the streamwise and spanwise directions as well as in time, are shown in Fig. \ref{Fig4}. The result trend is consistent with that of Gamahara \& Hattori~\cite{gamahara_searching_2017}, in which the same channel flow with $\text{Re}_{\tau} = 180$ was employed. The label data shows that the symmetric components ($\tau_{11}, \tau_{22}, \tau_{33}$) dominate owing to the high shear near the wall, which dissipates gradually farther from it. Here, $\tau_{12}$ reflects the momentum transfer in the wall-normal direction. Conversely, $\tau_{13}$ and $\tau_{23}$ exhibit values of approximately zero, thus contributing minimally to the SGS model. These variations in the values of the $\tau_{ij}$ components highlight their distinct significance in SGS modeling; additionally, they must be considered for determining the training setup for the $\tau_{ij}^{DNN}$ model.
    
    \subsection{\label{sec:level2}DNN algorithm and framework of data-driven SGS model}
    The purpose of the data-driven SGS model is to establish a functional relation between six output variables $\tau_{ij}$ and the information from resolved flow fields in the dataset. The appropriate input and output features must be selected to ensure that $\tau_{ij}^{DNN}$ model corresponds with its physical meaning. Similar to the conventional SGS model, the $\tau_{ij}^{DNN}$ model should be trained to satisfy the principle of material objectivity, as highlighted by Brenner et al.\cite{brener_highly_2024}, Wu et al.\cite{wu_physics-informed_2018}, Prakash et al.\cite{prakash_invariant_2022} and Ling et al.\cite{ling_reynolds_2016} Specifically, the $\tau_{ij}^{DNN}$ function should adhere to the Galilean and rotational invariant Navier--Stokes equation. To achieve this, $\bar{D}_{ij}$ is used as the input variable because it is Galilean and rotationally invariant. As mentioned by Brener et al.\cite{brener_highly_2024}, $\bar{D}_{ij}$ has a symmetric tensor, where each point of flow provides the quantity of an orthogonal basis associated with the unitary eigenvectors. Hence, accurate learning is enabled without depending on the reference frame. Additionally, $\bar{D}_{ij}$ incorporates $\frac{\partial \bar{u}_i}{\partial x_j}$, which includes the velocity and grid spacing. Since $\frac{\partial \bar{u}_i}{\partial x_j}$ is a function of the grid spacing, it can be consistently computed to obtain uniform grid-like information. It is expected to address the challenges of non-uniform grids in CNNs, as reported by Gao et al.\cite{gao_phygeonet_2021}, and maintains a consistent input for the $\tau_{ij}^{DNN}$ model across varying grid sizes. Here, the data-driven SGS model is expressed as  $\tau_{ij}^{DNN} = \mathcal{F}(X_{tr}, {w}) = \mathcal{F}(\bar{D}_{ij}, {w})$. The input ($X_{tr}$) and label ($S_{tr}$) feature for the training process can be defined as follows:
    
    \begin{equation} 
        X_{tr} = \{ x \in \mathbb{R}^{6 \times N_x \times N_y \times N_z} : x = \bar{D}_{ij} \}
        \label{input}
    \end{equation}  

    \begin{equation} 
        S_{tr} = \{ s \in \mathbb{R}^{6 \times N_x \times N_y \times N_z} : s = \tau_{ij}^{fDNS}\}
        \label{label}
    \end{equation}
    
   %
  
     The DNN architecture is critical, particularly for fluid dynamics problems, because it involves different scales that must be accurately captured. As mentioned by Liu et al.\cite{liu_investigation_2022} and Guan et al.\cite{guan_stable_2022}, CNN-based models consider non-local effects, thus reflecting the ability of DNN models to capture spatial dependencies and interactions instead of being limited to localized features. In this context, a multiscale SGS (MSC-SGS) model based on a CNN is proposed by incorporating the multiscale of turbulent vortices interactions and physical processes across scales.

    \begin{figure}
    \includegraphics[width=1\linewidth]{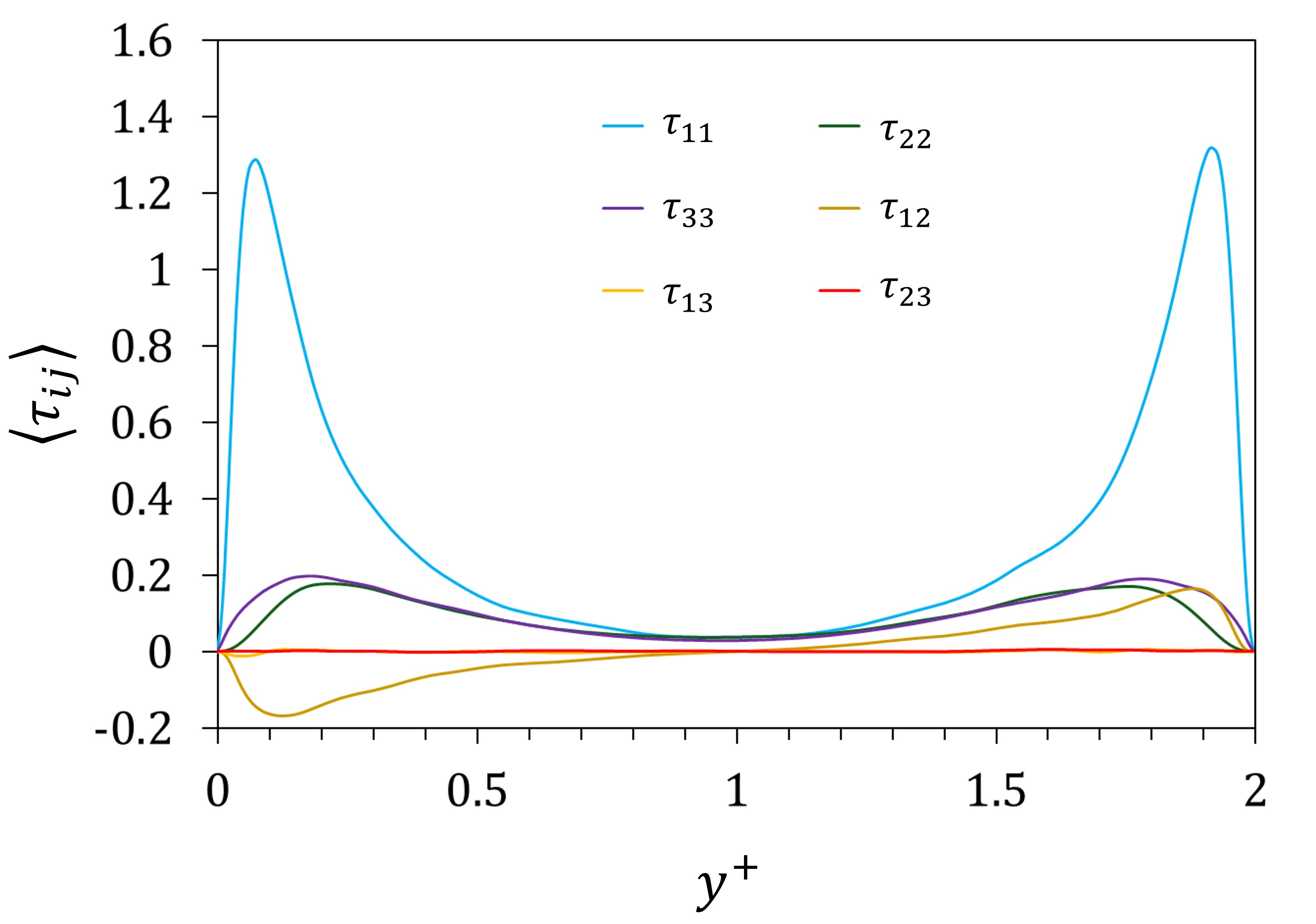}
    \caption{\label{Fig4} Wall normal distribution of $\tau_{ij}$ calculated from fDNS data. $\tau_{ij}$ is averaged in the streamwise-spanwise $x-z$ directions and in time.}
    \end{figure}
    
    \begin{figure}
    \includegraphics[width=1\linewidth]{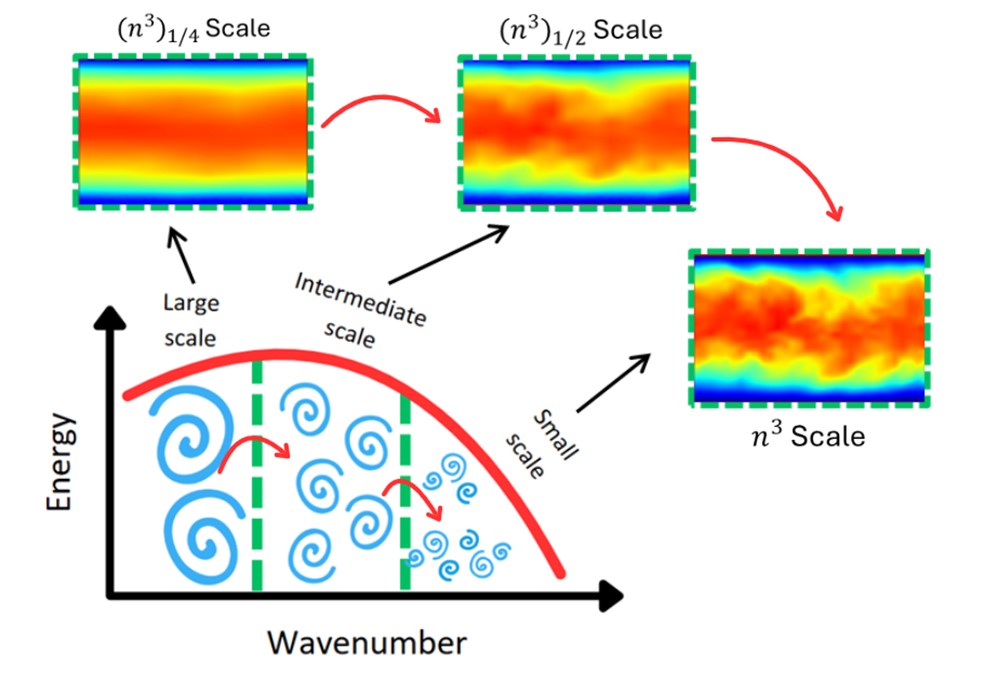}
    \caption{\label{Fig5} {Schematic diagram of the MSC-SGS model.}}
    \end{figure}
    
    \begin{figure}
    \includegraphics[width=1\linewidth]{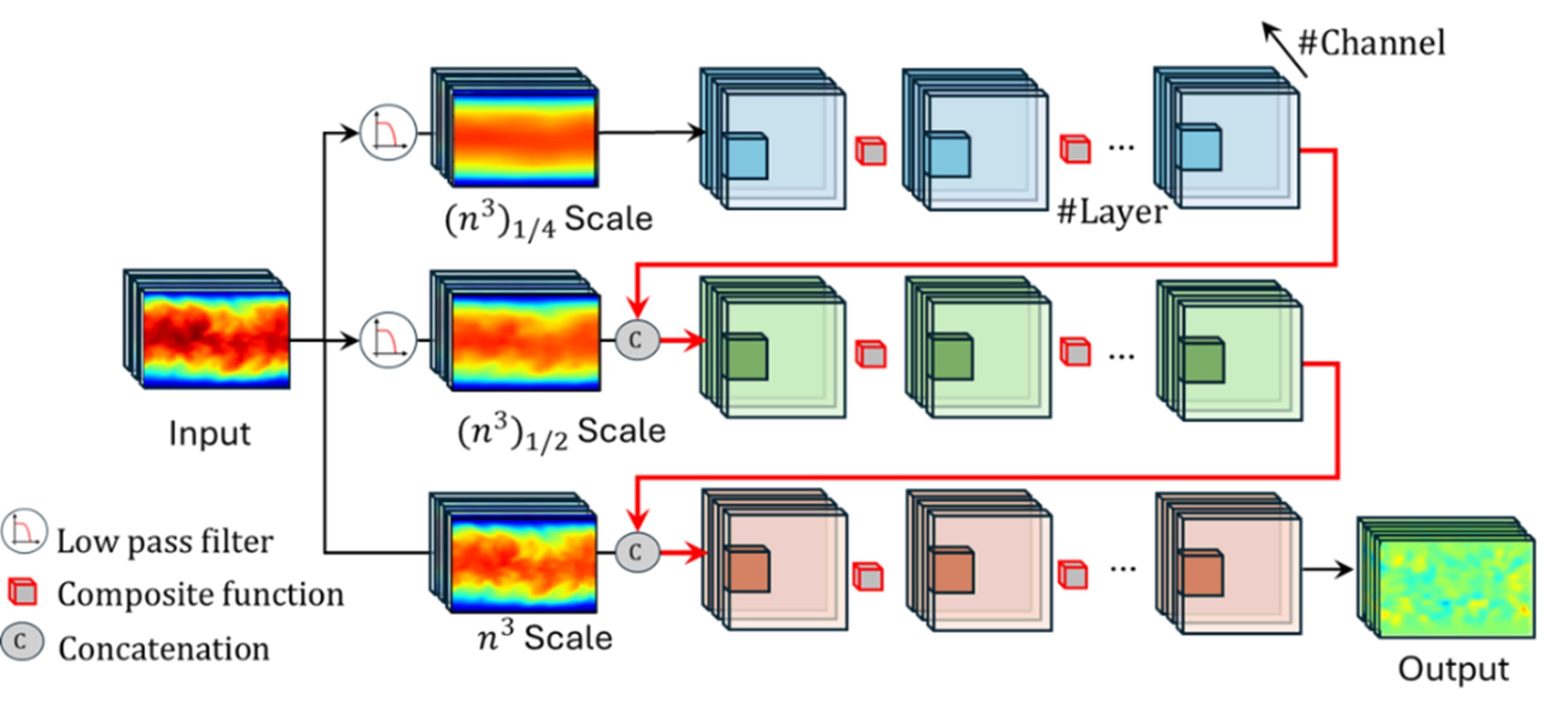}
    \caption{\label{Fig6} {Schematic illustration of MSC-SGS the network structure of MSC-SGS model.}}
    \end{figure}
    
    Figure \ref{Fig5} shows a schematic diagram of the concept of MSC-SGS model, which is based on the principle of energy cascades. To incorporate the flow of energy from a larger scale to a neighboring smaller scale into the DNN model, we consider incorporating representations that can address a wide range of scales. In this study, as shown in Fig. \ref{Fig6}, the flow fields at three scales—quarter, half, and full—obtained through low-pass filtering are fed into the MSC-SGS model as the input data. These scales correspond to large-, intermediate-, and scale-scale eddies, respectively, as illustrated in Fig. \ref{Fig5}. This step is crucial as it ensures that the multiscale model effectively captures the intricate details of energy distribution across scales. The multiscale algorithm shares conceptual similarities with multifidelity techniques\cite{mondal_multi-fidelity_2022,ajuria_illarramendi_performance_2022} to refine coarse-scale data into finer-scale representations. However, the algorithm used in our MSC-SGS model differs from that of the conventional multifidelity approach. Multifidelity approaches typically process each resolution using separate networks or models that are trained independently, thus potentially limiting the interaction between scales. By contrast, the MSC-SGS model integrates multiscale information within a single unified model. This model is expected to capture information across scales more effectively to address the limitations of conventional multifidelity techniques.

\begin{figure}
    \includegraphics[width=1\linewidth]{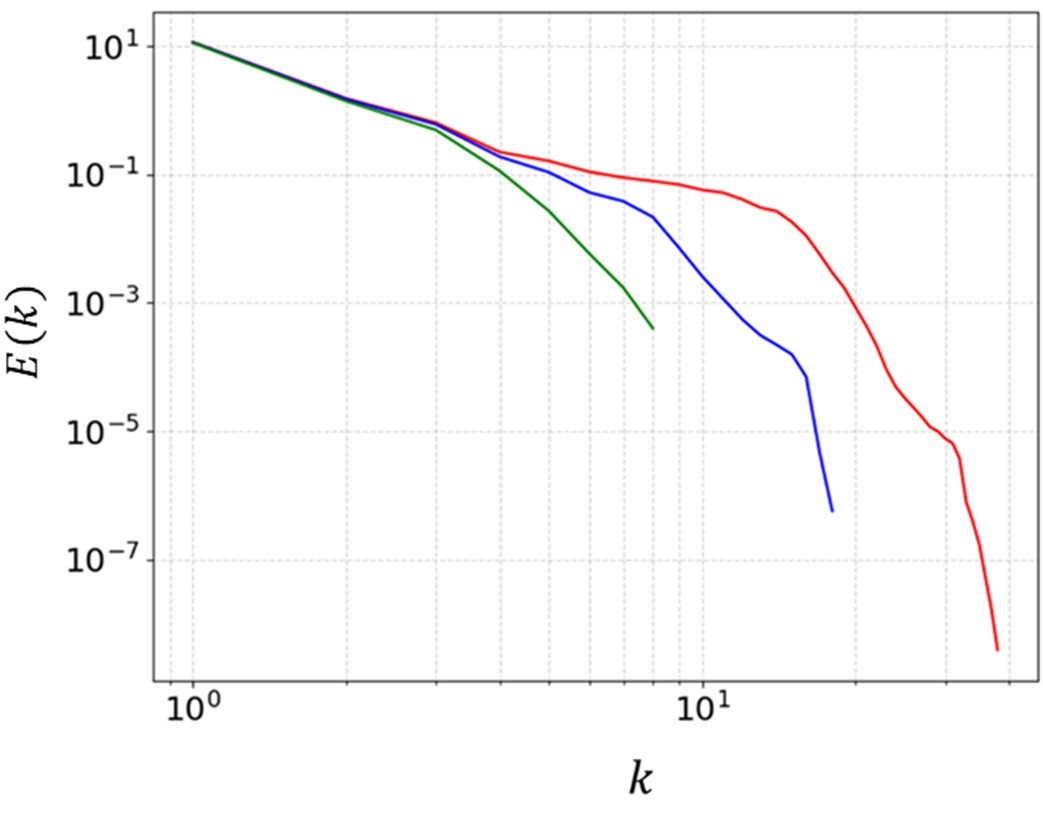}
    \caption{\label{Fig7} Energy spectrum with respect to different scales of MSC-SGS model. Red, blue, and green lines represent full scale, half scale, and quarter scale, respectively.}
    \end{figure}

    Figure \ref{Fig7} shows the energy spectra of the velocity field plotted for different input scales of the MSC-SGS model. The full scale retains the energy distribution across a broader range of wavenumbers, whereas the quarter scale predominantly captures the large-scale distribution. The half-scale and quarter-scale filter sizes indicate a systematic reduction, where the half- and quarter-scale filter measure $2\Delta$ and $4\Delta$ ($\Delta$: grid size), respectively, in each spatial direction. Each scale layer in the MSC-SGS model contains five hidden layers, which are determined through hyperparametric tuning. The MSC-SGS structure progressively encodes information, beginning from the largest scale and advancing to the smallest scale. Subsequently, the encoded information from each previous scale is concatenated to form the final multiscale representation. For example, the MSC-SGS model extracts information from the quarter-scale layer through a convolutional operation, which is then fed to the next layer as additional input. Based on hyperparameter tuning and energy spectrum analysis, we discover that using three scales is optimal. As mentioned by Huang et al.\cite{huang_densely_2018}, concatenation is suitable for combining the information between layers without diminishing the details, which can occur during the summation operation. This concatenation step represents an inter-scale cascade, as indicated by the red arrows in Fig. \ref{Fig5}.

    Now, we consider that input $(x_{k,l})$ passes through a convolutional layer, where $(.)_l$ denotes the hidden layer index $(l = 0 - 5)$ and $(.)_k$ the scale layer index, $k = 1,2$, and $3$ represent the quarter-, half-, and full-scale layers, respectively. The scale layers define the multiple input resolutions across different scales, whereas the hidden layers within each scale layer perform convolutional operations to encode information. Here, the composite function $( H_{k,l} )$ within the hidden layer comprises a convolutional layer, batch normalization (BN), and an activation function $(\varphi)$. More specifically, for the quarter layer, the output of each hidden layer is expressed as $x_{1,l} = H_{1,l}(x_{1,l-1})$. Subsequently, this encoded feature is passed to the half layer, where concatenation is performed such that $x_{2,1} = H_{2,1}([x_{2,0}, x_{1,5}])$, in which $[f, g]$ refers to the concatenation operation between $f$ and $g$. Similarly, the full-scale layer receives the encoded feature from the half-scale layer, as expressed by $x_{3,1} = H_{3,1}([x_{3,0}, x_{2,5}])$, which consists of the preceding values extracted from the quarter- and half-scale layers. This approach not only mimics the energy transfer between scales but also provides additional information similar to skip connections, thus preventing vanishing-gradient issues in deep networks.

    The convolutional layer computes feature maps by convolving the input data with a convolutional kernel (weight matrix) to detect spatial patterns. Considering the 3D spatial interactions in a turbulent flow, 3D kernels are employed in the convolutional layers with a uniform size of 3 in each spatial direction. To preserve the spatial information and address the boundary values, the same padding is applied\cite{raiaan_systematic_2024}. BN is introduced to suppress covariant shifts during training\cite{ioffe_batch_2015}. By normalizing the output of the convolutional layer, the extracted feature is adjusted to have a consistent mean of 0 and a standard deviation of 1. This normalization ensures the uniform scaling of features, thus improving stability and convergence. We set the batch size to 32, as obtained using the grid-search method\cite{raiaan_systematic_2024} in hyperparametric tuning. Next, to form the nonlinear function, the exponential linear unit (ELU) activation function ($\varphi$) is deployed. The ELU provides the advantage of a smoothing gradient during the back-propagation process by allowing negative values. This activation function is expected to be effective for extracting the flow field, such as the positive and negative components of ${\bar{D}_{ij}}$. As reported by Kim \& Lee\cite{kim_prediction_2020}, the ELU is more advantageous than the rectified linear unit (ReLU) because ReLU does not allow negative values, thus potentially eliminating the amount of information. In simple terms, the CNN layer can be expressed as shown in Eq.~\eqref{eq7}.
    
    \begin{equation}
    x_{k,l} = H_{k,l} \left( w * x_{k,l-1} \right) + b = \varphi \left( BN \left( w * x_{k,l-1} \right) \right) + b
    \label{eq7}
    \end{equation}
    where $w$ and $b$ are the convolutional kernel and bias, respectively; and $*$ denotes the convolutional operation.
    During training, optimization process is performed to determine the convolutional kernel ${w}$ that minimizes the loss function $L$, which are expressed as follows:
    
    \begin{equation}
        w = {\arg\min}_{{w}} \left( L(\tau_{ij}^{fDNS}, \tau_{ij}^{p}) \right)
        \label{eq8}
    \end{equation}
    
    \begin{equation}
        L = L_d + L_r = \sum_{i=1}^{3} \sum_{j=1}^{3} \gamma_{ij} \left(\tau_{ij}^{fDNS} - 	\tau_{ij}^{p} \right)^2 + \frac{\lambda}{2} \sum_{k,l} w_{k,l}^{2}
        \label{eq:loss}
    \end{equation}
    
    The loss function $(L)$ comprises the prediction loss $(L_d)$ and regularization loss $(L_r)$, as expressed in Eq.~\eqref{eq:loss}. As shown in Fig. \ref{Fig4}, the averaged values of $\tau_{13}$ and $\tau_{23}$ are approximately zero, which renders it difficult for the $\tau_{ij}^{DNN}$ model to update $w$ and construct a reliable nonlinear regression function. In fact, this issue has been reported by Gamahara \& Hattori\cite{gamahara_searching_2017}, Kang et al.\cite{kang_neural-network-based_2023}, and Bose \& Roy\cite{bose_invariance_2024}, where a low correlation was observed between the predicted components and fDNS data. Therefore, we introduce a weighting parameter $\gamma$ to magnify the effect of $\tau_{ij}$, thus facilitating a more straightforward update for ${w}$. We select $\gamma$ based on the global average value of $\tau_{ij}$ components and through a hyperparametric step. The values of $\gamma_{ij}$ were $\gamma_{11} = 1$, $\gamma_{22} = 10$, $\gamma_{33} = 10$, $\gamma_{12} = 10$, $\gamma_{13} = 10$, and $\gamma_{23} = 100$. Additionally, we add a regularization loss $(L_r)$ to prevent overfitting with the regularization constant $\lambda$, which is set to $10^{-4}$. Here, ${w}$ denotes the kernel coefficient of the convolutional layers and $(.)_{k,l}$ indicates the scale and hidden layer index, respectively.
    
    To minimize loss, we use the gradient descent method of adaptive moment estimation (ADAM)\cite{kingma_adam_2017}, which iteratively updates the values of the kernel and bias in the negative direction of the total loss gradient. Moreover, we apply the scheduler learning rate, whose initial value was $10^{-3}$ and reduced by $1/10$ at every $100$ epochs. Here, an epoch refers to the period in which the training datasets are used simultaneously for training. For the MSC-SGS model code, we adopt the algorithm of Illarramendi et al\cite{ajuria_illarramendi_performance_2022}, written in Python. The training is conducted using the machine-learning library of PyTorch 1.13.1, and the computation is executed on a large-scale computer system comprising an 8 GPU core on an NVIDIA A100 GPU at the D3 Center of Osaka University.
    
    \subsection{\label{sec:level2}Target of comparison}
\begin{figure}[t]
    \centering
        \begin{subfigure}{0.4\textwidth}
            \centering
            \includegraphics[width=\linewidth]{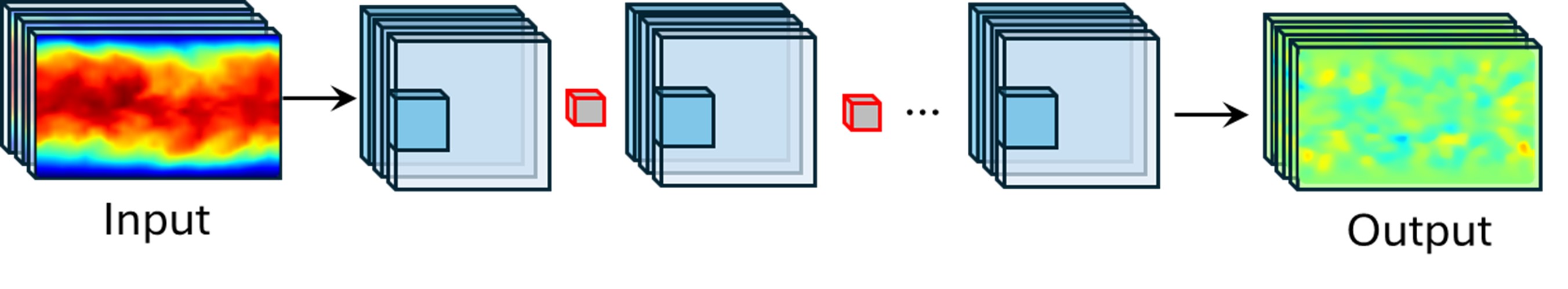}
            \caption{}
            \label{Fig8a}
        \end{subfigure}
        \hfill
        \begin{subfigure}{0.4\textwidth}
            \centering
            \includegraphics[width=\linewidth]{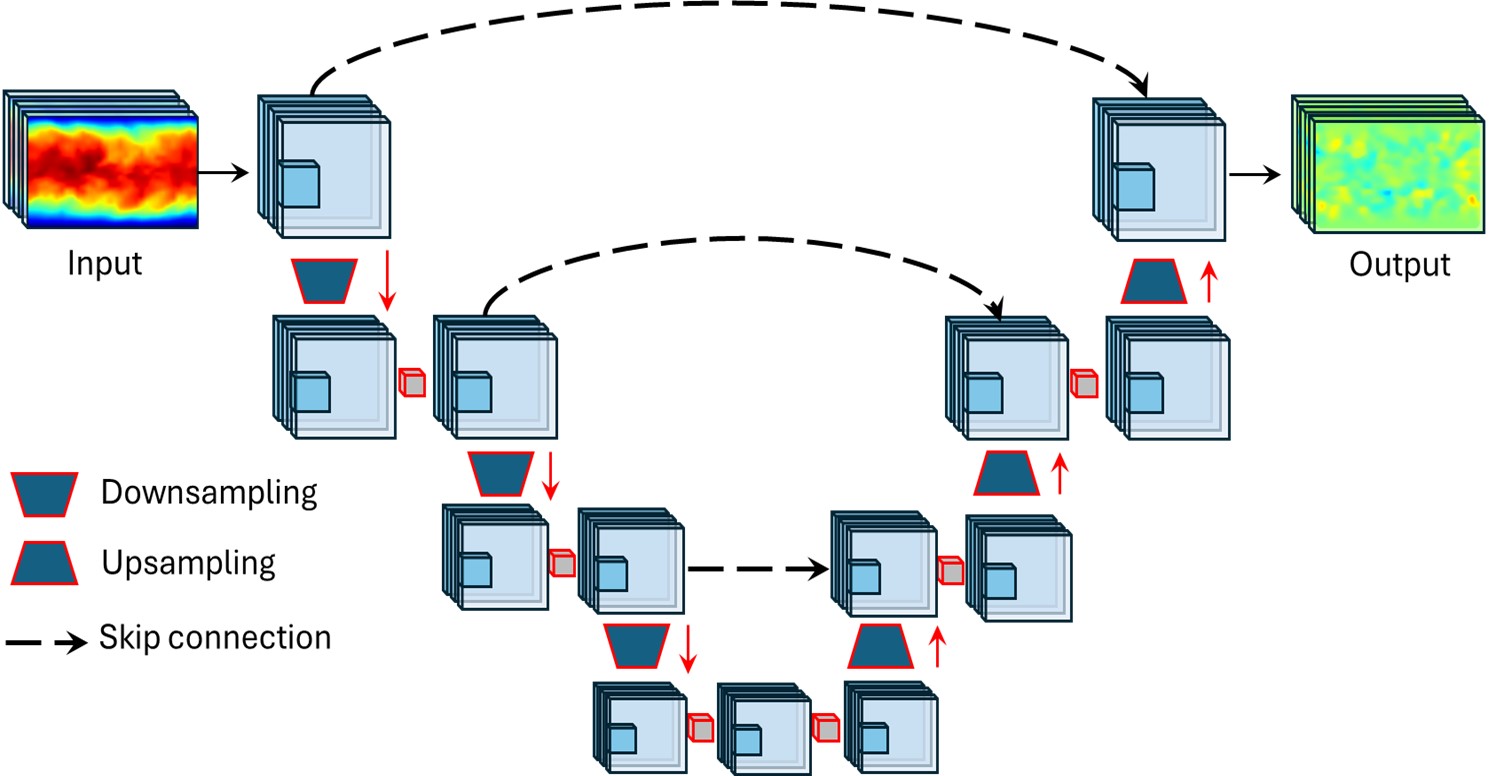}
            \caption{}
            \label{Fig8b}
        \end{subfigure}
        \caption{Schematic illustrations of (a) monoscale algorithm and (b) U-Net algorithm.}
        \label{Fig8}
    \end{figure}

    To evaluate the performance of the MSC-SGS model, a comparative study is conducted using two CNN architectures: the monoscale CNN and the U-Net. The comparative analysis is performed to assess the effects of the different architectures on the ability to capture turbulence field features. The simplified algorithms are illustrated in Fig. \ref{Fig8}. Both models are trained using similar training setups and datasets to ensure a consistent evaluation. A 3D convolutional kernel is applied to both models. The monoscale model comprises 10 sequential layers optimized through hyperparameter tuning. The same padding is applied to preserve the spatial dimensions throughout the network. The $\bar{D}_{ij}$ of the full scale flow field is fed as the input. This model focuses on extracting features without explicitly accounting for multiscale representation. It relies merely on the layer depth, as demonstrated by Liu et al.\cite{liu_investigation_2022} and Guan et al.\cite{guan_stable_2022}
    
    By contrast, the U-Net model is designed to capture multiscale spatial features by employing an encoder–decoder structure with a progressive downsampling and upsampling structure\cite{ajuria_illarramendi_performance_2022}. In this study, average pooling is applied for downsampling, whereas nearest-neighbor interpolation is employed for upsampling to preserve the spatial accuracy. Saura \& Gomez\cite{saura_predicting_2023} also employed a data-driven SGS model based on the U-Net. The input is initially fed into the network at full resolution, which progressively reduces the spatial resolution through a downsampling operation into a lower-dimensional manifold. Subsequently, the upsampling operation restores the resolution, with skip connections transfers high-resolution features directly from the encoder to the decoder layers. These skip connections are introduced to mitigate vanishing-gradient issues associated with deep networks and ensure the preservation of fine-scale information.

\section{\textit{A priori} test results}
In \textit{a priori} test, we examine the accuracy of the $\tau_{ij}^{DNN}$ model in predicting $\tau_{ij}$ with unseen data from the testing dataset. The performance of the SGS model is evaluated by comparing the physical quantities of $\tau_{ij}$ using the correlation coefficient (CC) expressed in Eq.\eqref{eq10}. Here, $\langle . \rangle$ denotes the ensemble average in the streamwise–spanwise ($x$–$z$) direction and in time.

\begin{equation}
    CC= \frac{\langle (\tau_{ij}^{fDNS}-\langle \tau_{ij}^{fDNS} \rangle)(\tau_{ij}^{p}-\langle \tau_{ij}^{p} \rangle)\rangle}
    {\sqrt{\langle (\tau_{ij}^{fDNS}-\langle \tau_{ij}^{fDNS} \rangle)^2 \rangle } \sqrt{\langle (\tau_{ij}^{p}-\langle \tau_{ij}^{p} \rangle)^2 \rangle }}
    \label{eq10}
\end{equation}

\begin{figure}
    \includegraphics[width=1\linewidth]{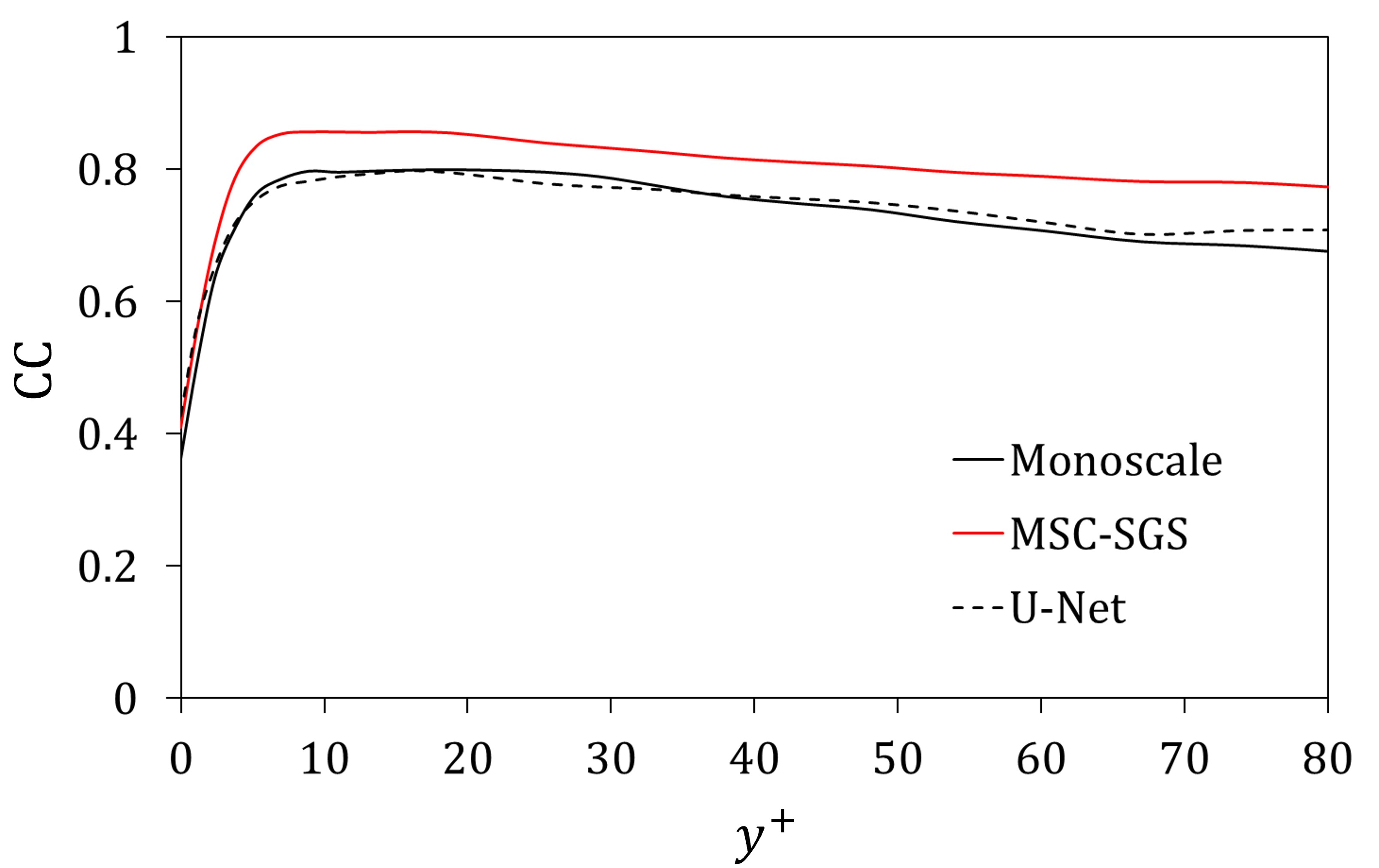}
    \caption{\label{Fig9} Wall-normal distributions of correlation coefficients (CC) averaged in streamwise–spanwise ($x-z$) direction and in time between $\tau_{ij}^{fDNS}$ and $\tau_{ij}^p$.}
    \end{figure}

Figure \ref{Fig9} shows the CC plotted as a function of wall distance ($y^+$) averaged for all six components of $\tau_{ij}$. The MSC-SGS model consistently achieves the highest CC within the $y^+$ range. Within $y^+ < 30$, the MSC-SGS model provides a high correlation, where shear-dominated turbulence and SGS dynamics are the most active. In the viscous sublayer region, the correlations of all models are lower because this region is characterized by dominant viscous stress and weak turbulent fluctuations. The limited variability in the flow features renders it challenging for the $\tau_{ij}^{DNN}$ model to extract information. This result is reasonable and consistent with the findings of Gamahara \& Hattori\cite{gamahara_searching_2017}. For $y^+ > 30$, all models depict slightly deteriorated performances as the SGS dynamics become less dominant. This can be attributed to the homogeneity in the far wall, which limits the distinction between multiscale features. However, the MSC-SGS model outperforms the monoscale and U-Net models, thus highlighting its ability to capture flow-field information in regions where the structure is more homogeneous.

Table \ref{table:CC} shows the average CC of each $\tau_{ij}$ component, computed over the entire domain and compared between data-driven models. As shown, the monoscale and U-Net models attain similar CC, whereas the MSC-SGS model outperforms both by approximately 5.5\% on average. The component of $\tau_{11}$ exhibits the highest CC for all models owing to its high and dominant value, which suggests that it is easier to be predicted. For lower values of $\tau_{13}$ and $\tau_{23}$, all models provide reasonable results, thus demonstrating that applying the present loss function (Eq.\eqref{eq:loss}) enhances the ability of the DNN model to effectively predict these components. For comparison, the CC results yield higher values compared with those obtained by previously reported models, such as those of Gamahara \& Hattori\cite{gamahara_searching_2017}, Park \& Choi\cite{park_toward_2021}, and Bose \& Roy\cite{bose_invariance_2024}. The MSC-SGS model exhibits greater improvements over other models, with accuracy gains of approximately 7.8\% and 9.4\% for $\tau_{13}$ and $\tau_{23}$, respectively. The difference in the DNN architecture demonstrates the significance of the multiscale approach via the higher accuracy afforded. Compared with the MSC-SGS model, the U-Net lacks explicit multiscale inputs, which may limit its capacity to capture intricate scales despite its deep architecture. Similarly, the monoscale model struggles to capture fine-scale structures because of its inherent architectural limitations.

\begin{table*}[ht]
    \centering
    \small 
    \renewcommand{\arraystretch}{1.5}
    \caption{CC between $\tau_{ij}^{fDNS}$ and $\tau_{ij}^{p}$ for six $\tau_{ij}$ components. $\overline{CC}$ denotes average CC in all six components of $\tau_{ij}$.}
    \begin{tabular}{lccccccc}
        \toprule
        \text{$\tau_{ij}^{DNN}$ model} & $\tau_{11}$ & $\tau_{22}$ & $\tau_{33}$ & $\tau_{12}$ & $\tau_{13}$ & $\tau_{23}$ & $\overline{CC}$ \\
        \hline
        MSC-SGS & 0.926 & 0.844 & 0.865 & 0.889 & 0.799 & 0.725 & 0.841 \\
        Monoscale & 0.902 & 0.802 & 0.827 & 0.851 & 0.741 & 0.655 & 0.797 \\
        U-Net & 0.893 & 0.799 & 0.832 & 0.847 & 0.737 & 0.657 & 0.794 \\
        \toprule
    \end{tabular}
    \label{table:CC}
\end{table*}

\begin{figure*}[ht]
    \centering
    \subfloat[\label{Fig10a}]{\includegraphics[width=0.45\linewidth]{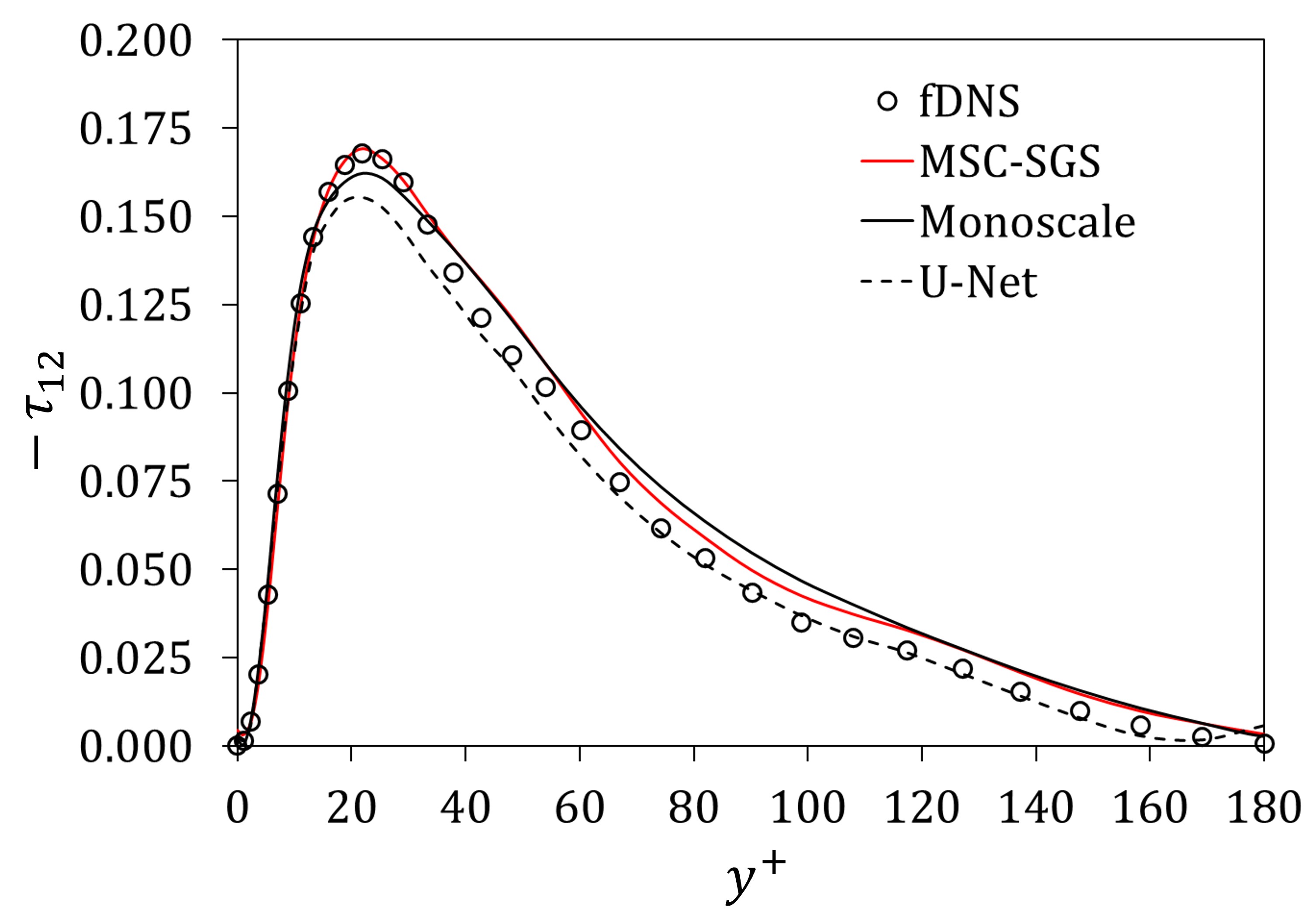}}\hspace{0.5cm}
    \subfloat[\label{Fig10b}]{\includegraphics[width=0.45\linewidth]{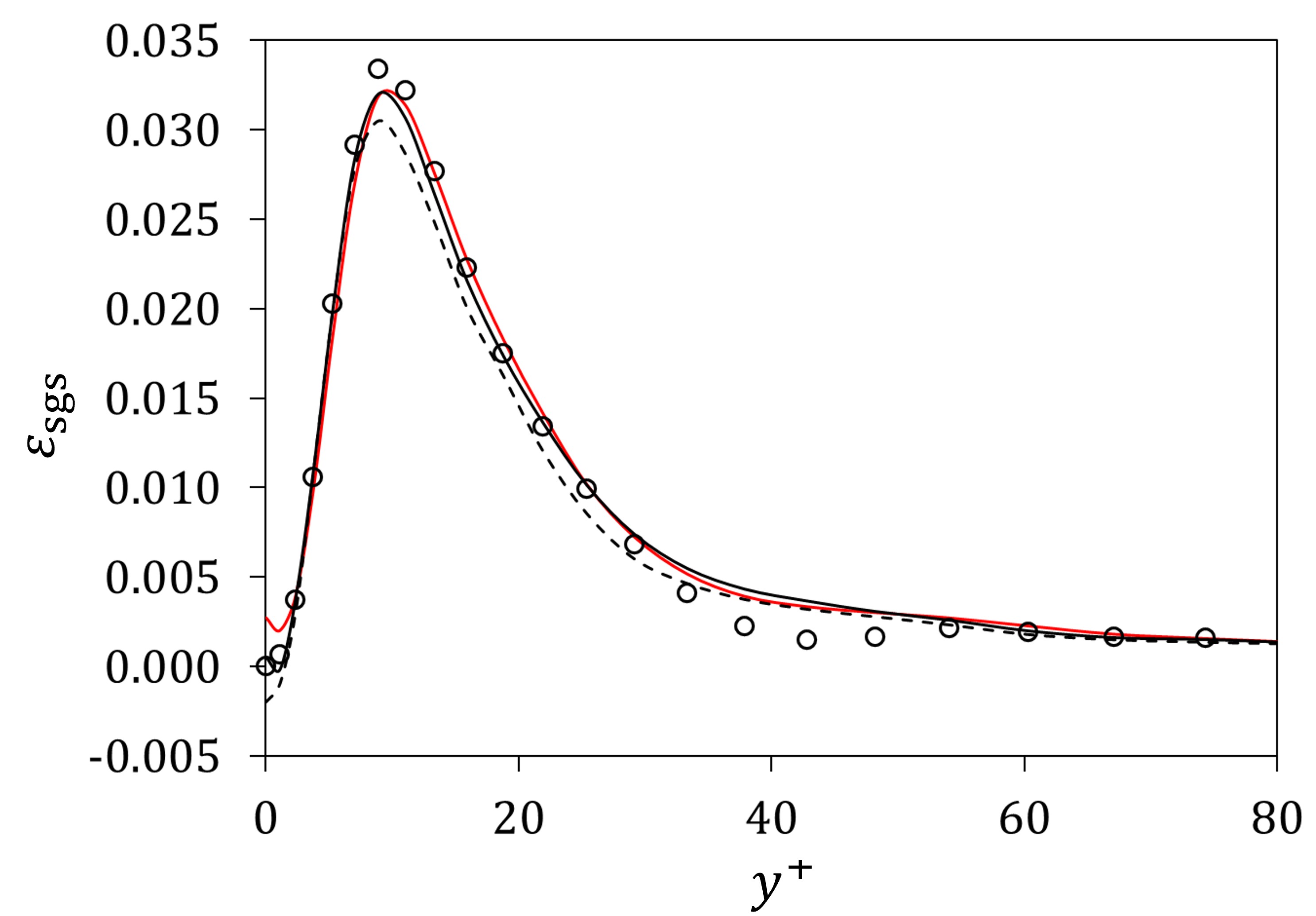}}\\
    \subfloat[\label{Fig10c}]{\includegraphics[width=0.45\linewidth]{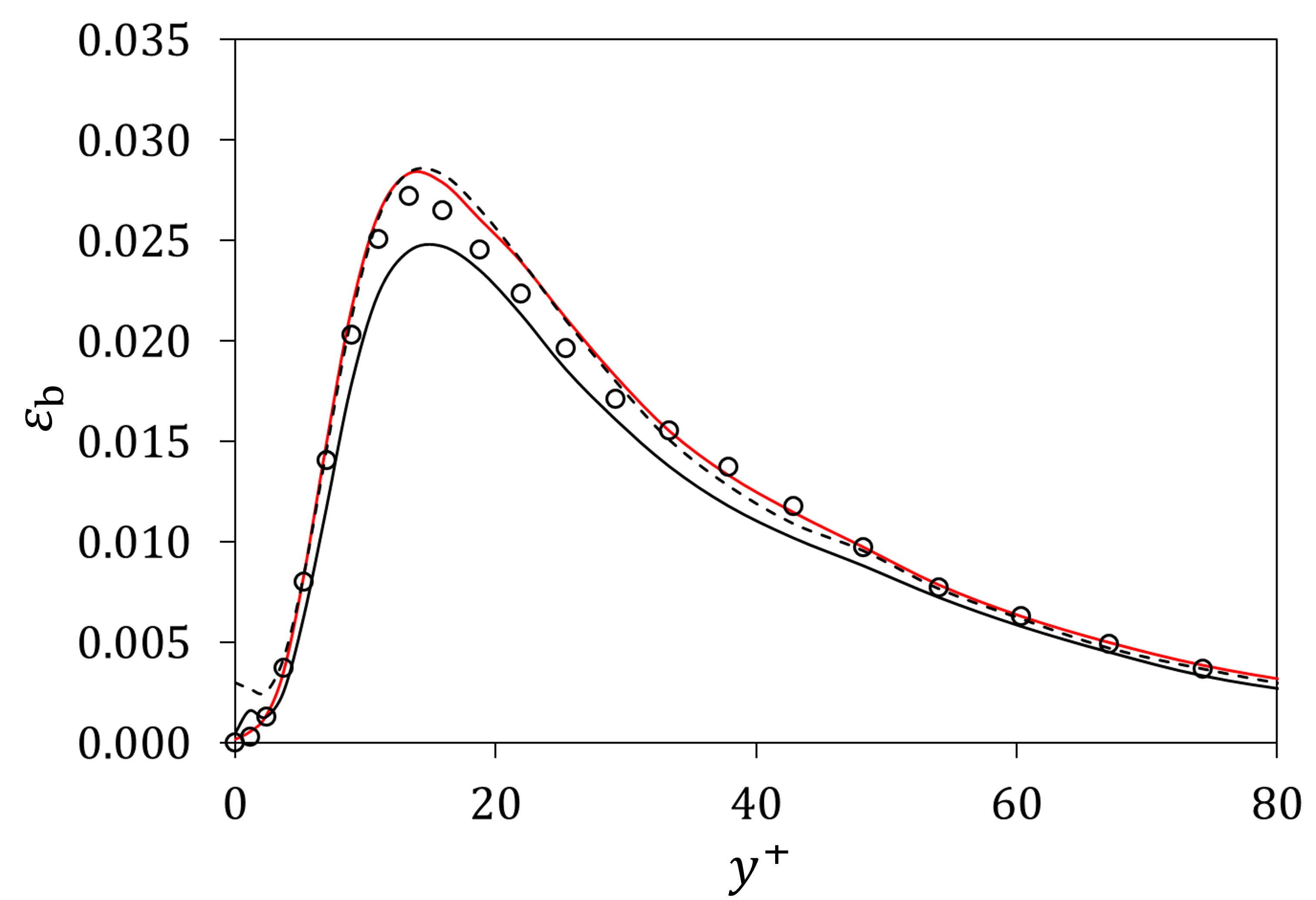}}\hspace{0.5cm}
    \subfloat[\label{Fig10d}]{\includegraphics[width=0.45\linewidth]{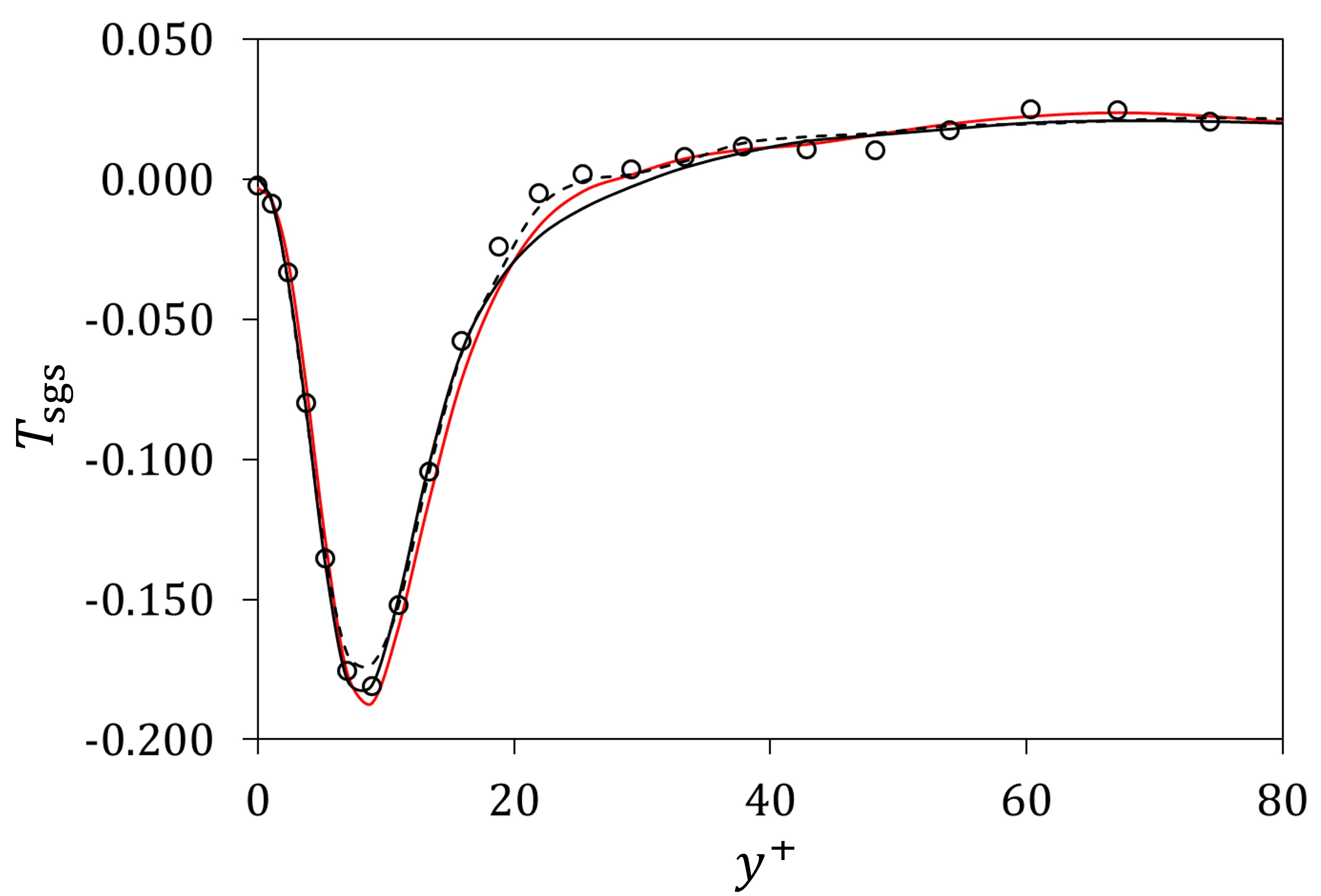}}
    
    \caption{Wall-normal distributions of (a) $\tau_{12}$, (b) SGS dissipation, (c) SGS backscatter, and (d) SGS transport.}
    \label{Fig10}
\end{figure*}

Figure \ref{Fig10}(\subref{Fig10a}) shows the wall-normal distribution of $\tau_{12}$, which is the antisymmetric shear-stress component. Among the six components of $\tau_{ij}$, $\tau_{12}$ is considered the most important component in wall-bounded flow\cite{bose_invariance_2024}. Fig. \ref{Fig10}(\subref{Fig10b}) illustrates the SGS dissipation under the assumption of local equilibrium, defined as $\varepsilon_{sgs}=-\tau_{ij} \bar{D}_{ij}$. In both $\tau_{12}$ and $\varepsilon_{sgs}$, the MSC-SGS and monoscale models agree well with the fDNS data. Although the U-Net underestimates the peak values, the peak positions and spatial distribution trends are accurate. For the region $y^+ \approx 40$ shown in Fig. \ref{Fig10}(\subref{Fig10b}), all models show slight increases, which are attributable to inaccuracies in capturing information during the transition from small-scale turbulence, which is dominated by viscous forces, to large-scale turbulence, which is governed by inertial forces. In the MLP model of Park \& Choi\cite{park_toward_2021}, there were misalignment of peak positions, over- or underestimations, and differences in the distribution trends. This indicates that incorporating neighbor information as input to the CNN-based model contributes positively to the accuracy. In fact, this was confirmed by Park and Choi\cite{park_toward_2021}, where the prediction accuracy improved when the variables of multiple grid points were used as input compared with the case when single grid points were used as input.

Figure \ref{Fig10}(\subref{Fig10c}) presents the SGS backscatter ($\varepsilon_b=-\frac{1}{2} \langle \varepsilon_{sgs}-|\varepsilon_{sgs}| \rangle$), which shows the reverse transfer of energy from SGSs to resolved scales. This phenomenon is important for wall-bounded turbulence because it is related to turbulence phenomena such as bursting events\cite{park_toward_2021}. The monoscale model underpredicts the backscatter by a mean deviation of 1.89\%. By contrast, both the MSC-SGS and U-Net models yield predictions that align more closely with the fDNS results. This implies that treating the interactions between various scales promotes energy cascades. The MLP-based model$^8$ showed a lower $\varepsilon_b$ value than the present MSC-SGS model in \textit{a priori} testing. This indicates that CNN-based models are effective for predicting the inverse cascade to capture the spatial distribution and interactions of variables. In general, models that express inverse cascades in LES tend to make the computational unstable. Hence, the computation with MSC-SGS model is expected to be numerically unstable in \textit{a posteriori} tests compared with other models, such as that of Park \& Choi\cite{park_toward_2021} and the conventional Smagorinsky (SMAG) model. The SGS transport ($T_{sgs}=\frac{\partial (\tau_{12} \bar{u}_i)}{\partial x_j}$), as shown in Fig. \ref{Fig10}(\subref{Fig10d}), is examined to accurately represent energy transfer in the LES\cite{volker_optimal_2002}. All three CNN-based models showed excellent agreement with the fDNS data, with prediction accuracies better than those afforded by the MLP-based models\cite{park_toward_2021,bose_invariance_2024}. A precise $T_{sgs}$ representation ensures that the energy flux between the resolved scale and SGS is balanced, thus resulting in improved numerical stability.

\begin{figure}
\includegraphics[width=1\linewidth]{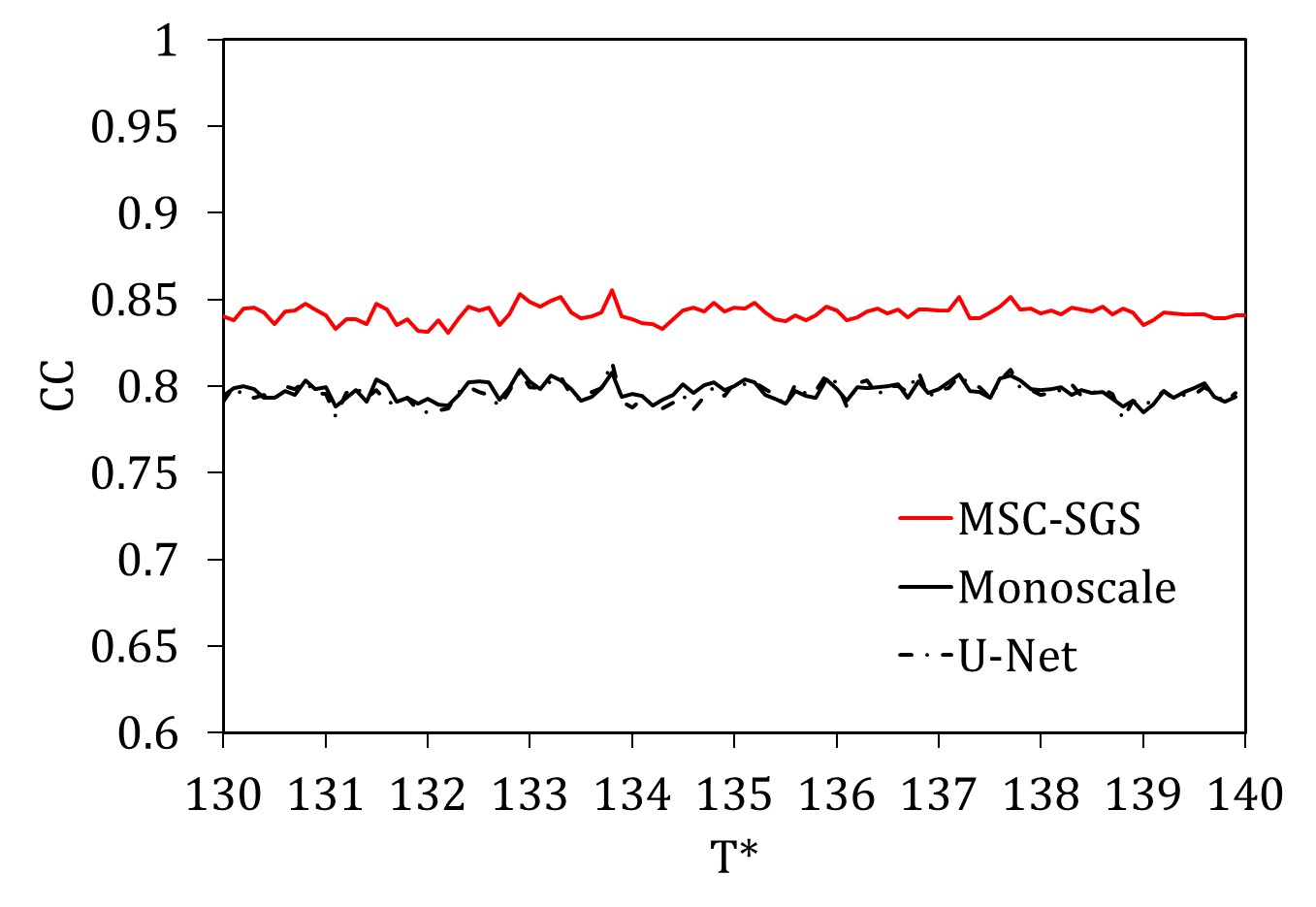}
\caption{\label{Fig11} {Temporal evolution of averaged CC computed over entire domain.}}
\end{figure}

\begin{figure}
\includegraphics[width=1\linewidth]{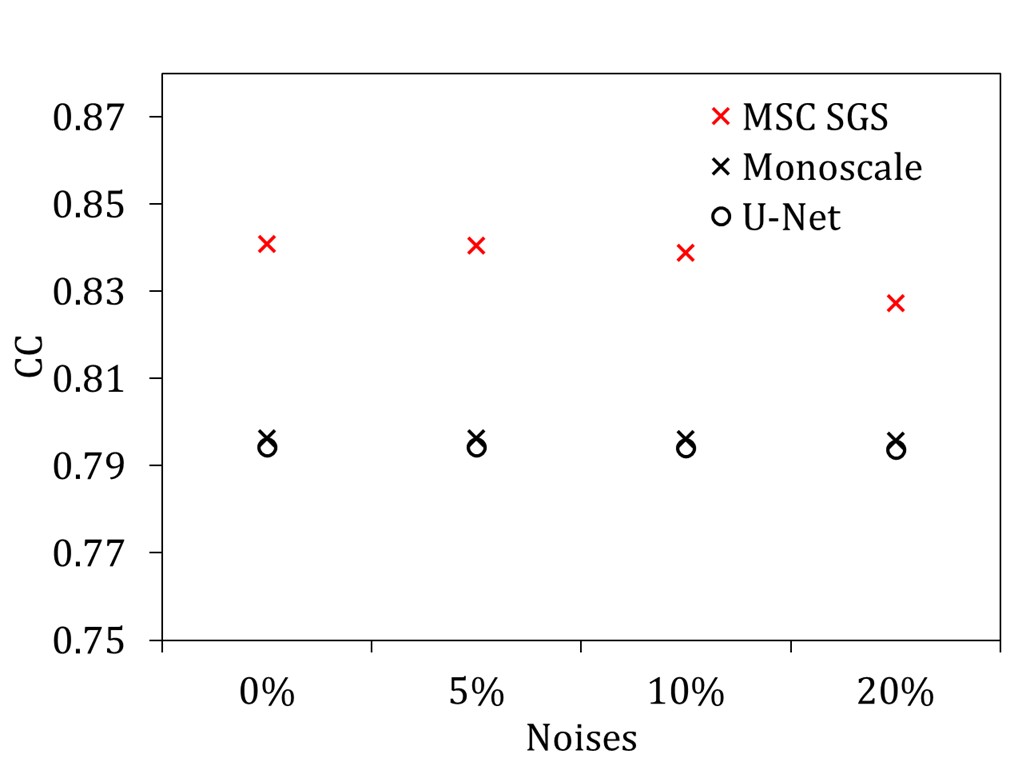}
\caption{\label{Fig12} {Domain- and time-averaged CC values with respect to noise levels.}}
\end{figure}

To assess the stability and robustness of the DNN models, we evaluate the temporal accuracy of the CC by introducing Gaussian noise into the input data. This approach measures the sensitivity of the DNN model under perturbed or invalid input conditions, as highlighted by Goodfellow et al.\cite{goodfellow_deep_2016} and Kim et al.\cite{kim_large_2024} Here, the inputs are changed such that $\bar{D}'_{ij} = \bar{D}_{ij} + N(0, \sigma_{ij}^2)$, where $N$ represents a zero-mean normal distribution with a standard deviation of $\sigma$. Fig. \ref{Fig11} shows the temporal evolution of the average CC value in the entire domain for all models over a range of instantaneous times ($\text{T$^*$} = t u_{\tau}^2 / \nu$) for a 5\% noise level. Additionally, 10\% and 20\% noise levels are introduced. Fig. \ref{Fig12} shows the time-averaged CC, which shows that even though the noise level increases, all the DNN models consistently maintain a steady CC. The inclusion of regularization-loss attributes prevented overfitting during the training process, thus ensuring that the model generalized to temporal and noisy input data. Although this method provides insights into the robustness of the model, it does not fully capture the actual performance in representing turbulent flow, which is affected significantly by vortex structures and their dynamic behavior. Despite these limitations, this assessment serves as a baseline for demonstrating the sensitivity of the model when exposed to incorrect inputs.

In \textit{a priori} tests, CNN-based models generally show better agreement with reference fDNS data as compared with MLP-based models. This indicates the effectiveness of CNN-based models in capturing the spatial distribution and interactions of variables. However, as reported in previous studies (Maulik et al.\cite{maulik_data-driven_2018} Park \& Choi\cite{park_toward_2021}, Guan et al.\cite{guan_stable_2022}, and Duraissamy\cite{duraisamy_perspectives_2021}), \textit{a priori} results do not guarantee accuracy or stability in \textit{a posteriori} tests. \textit{A priori} analysis evaluates models using data by assuming perfect conditions while disregarding the effects of numerical errors and the temporal evolution of uncertainties in the actual flow. By contrast, \textit{a posteriori} tests incorporate these complexities, where data-driven SGS models are implemented within LES frameworks and their predictive performance is evaluated under dynamic and realistic flow conditions.

\section{\textit{A posteriori} test results}
\begin{figure}
\includegraphics[width=1\linewidth]{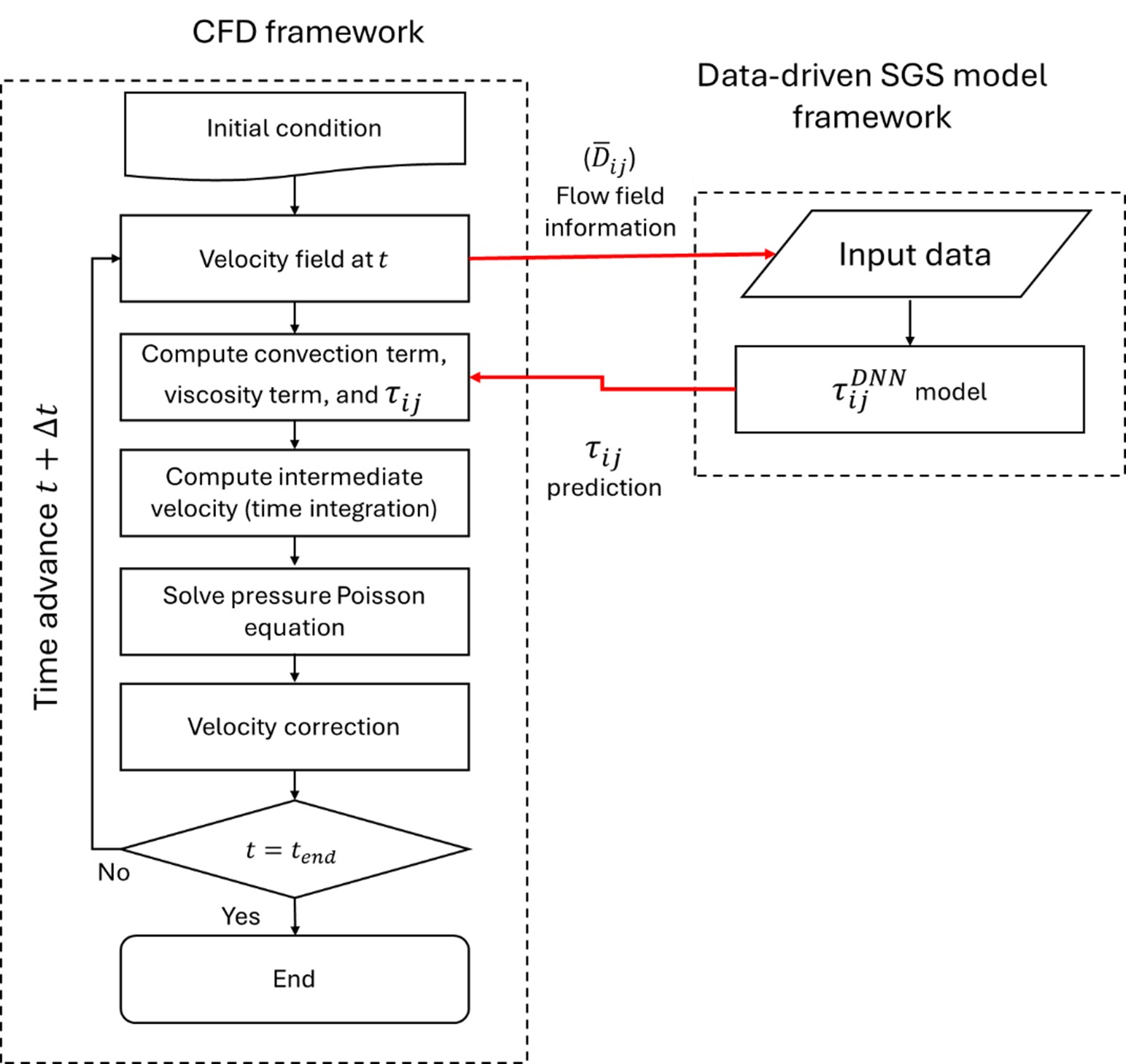}
\caption{\label{Fig13} {Schematic diagram of CFD and data-driven SGS model framework.}}
\end{figure}

In \textit{a posteriori} test, the data-driven SGS model is implemented in an actual LES with a problem setting of turbulent channel flow similar to that of the DNS. The LES governing equations are expressed as shown in Eqs. \eqref{mass} and \eqref{momentum}. The numerical methods are the same as those of the DNS, while the domain size and grid points are $(L_x \times L_y \times L_z) = (2\pi\delta \times 2\delta \times \pi\delta)$ and $(n_x \times n_y \times n_z) = (32 \times 64 \times 32)$, respectively. To address negative dissipation ($\tau_{ij}^{p} \bar{D}_{ij} < 0$), clipping is applied to stabilize the computation, which is similarly conducted by Maulik et al.\cite{maulik_data-driven_2018} and Park \& Choi\cite{park_toward_2021}. However, this artificial manipulation is neglected in the present LES as the occurrence rate of negative dissipation is zero, and computations with all $\tau_{ij}^{DNN}$ models are numerically stable. For comparison, the SMAG model with the van Driest damping function is used as a conventional SGS model. \textit{A posteriori} test framework is shown in Fig. \ref{Fig13}. In the conventional LES, $\tau_{ij}$ is computed within the CFD framework shown in Fig. \ref{Fig13}. In a data-driven LES, the flow-field information ($\bar{D}_{ij}$) is passed to the data-driven SGS model framework. It processes the flow-field information as input and predicts the $\tau_{ij}$ value. Subsequently, the $\tau_{ij}$ values are re-integrated into the CFD framework to update the flow field.

A mixed-language programming approach is used in the \textit{a posteriori} test, i.e., Python for the data-driven SGS model framework and Fortran for the CFD framework. Fortran is selected because of its superior computational efficiency compared with Python, thus ensuring a faster simulation. To integrate the process, shell commands are utilized to execute Python scripts within Fortran, which allowed the $\tau_{ij}^{DNN}$ model to be integrated seamlessly into the CFD framework. Computations for the \textit{a posteriori} test are conducted using a large-scale computer system comprising 16 cores on an Intel Xeon Platinum CPU at the D3 Center at Osaka University.

\begin{figure*}[ht]
    \centering
    \subfloat[\label{Fig14a}]{\includegraphics[width=0.45\linewidth]{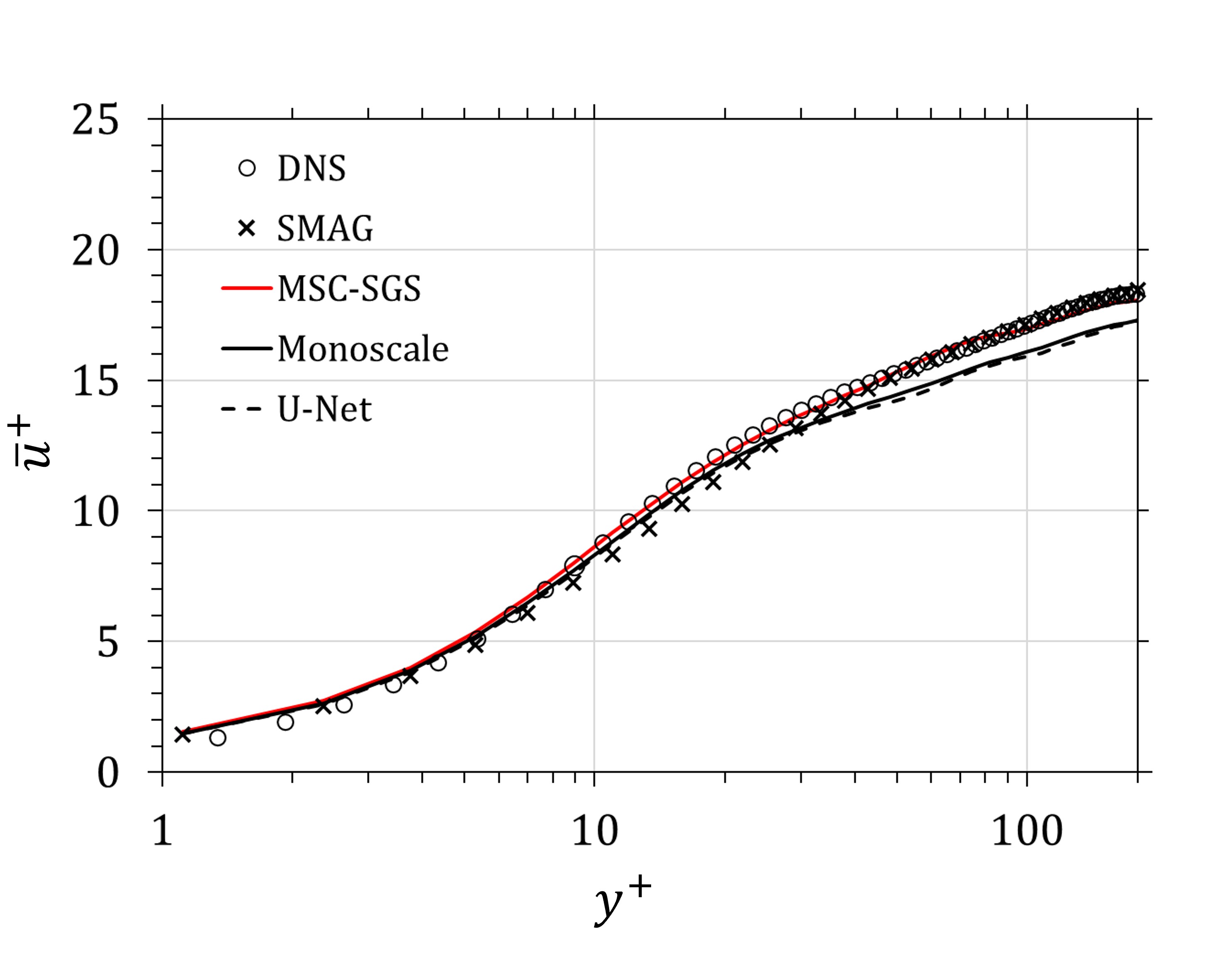}}\hspace{0.2cm}
    \subfloat[\label{Fig14b}]{\includegraphics[width=0.45\linewidth]{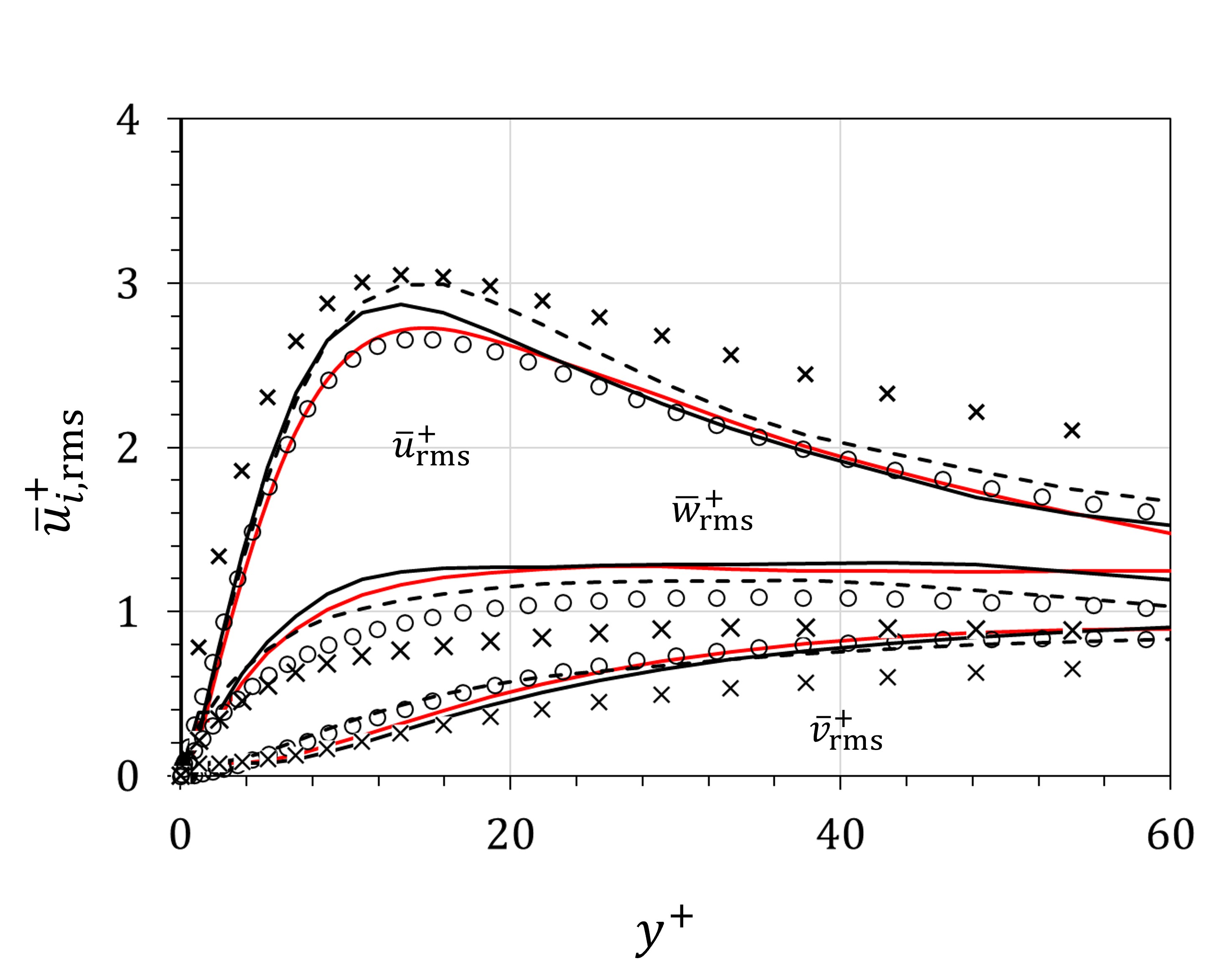}}\\
    \subfloat[\label{Fig14c}]{\includegraphics[width=0.45\linewidth]{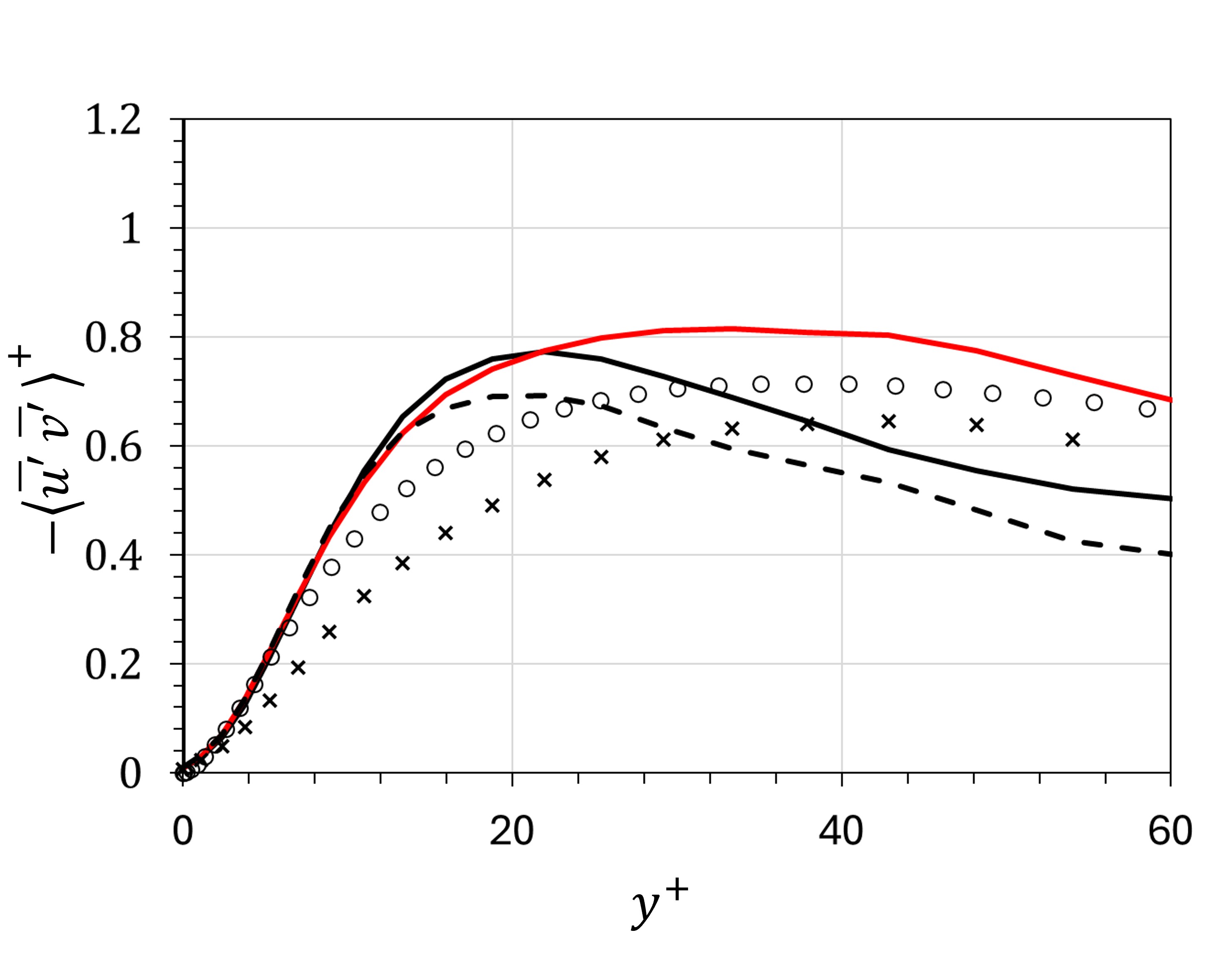}}\hspace{0.2cm}

    \caption{Wall-normal distribution of turbulence statistics of \textit{a posteriori} result for (a) mean velocity, (b) root-mean-square velocity, and (c) Reynolds stress.}
    \label{Fig14}
\end{figure*}

Figure \ref{Fig14}(\subref{Fig14a}) shows a comparison of the mean velocity profiles of the models and reference DNS. In general, all models successfully capture the overall trend of the DNS data. In the $y^+ < 50$ region, where viscous effects dominate and turbulence is anisotropic, all SGS models exhibit reasonable agreement with the DNS data. However, slight discrepancies emerge because of the challenges in resolving fine-scale turbulent structures near the wall. In the $5 < y^+ < 40$ region, the SMAG shows significant deviations while the $\tau_{ij}^{DNN}$ models profile are considerably consistent with the DNS data. The SMAG uses the eddy viscosity model coupled with the van Driest damping function, which results in deviations in the transitional region of the wall and logarithmic layer. By contrast, the $\tau_{ij}^{DNN}$ models trained on the fDNS datasets achieve better mean velocity profiles by directly capturing flow dynamics without relying on the eddy viscosity model and damping assumptions. In the $y^+ > 50$ region, where turbulence become more isotropic, the MSC-SGS model and SMAG outperform the monoscale and U-Net models by maintaining a closer match with the DNS results. Conversely, the monoscale and U-Net models exhibit clear deviations, with overall mean errors of approximately 8.1\% and 7.7\%, respectively. These deviations may be attributed to accumulation error arising from inaccuracies in predicting $\tau_{ij}$, which suppress the mean velocity. As highlighted by Durraissamy$^{35}$, in \textit{a posteriori} test, the model relied on coarse-grained features different from the fDNS data, thus resulting in an overestimated prediction for $\tau_{ij}$ that progressively accumulated over time.

Figure \ref{Fig14}(\subref{Fig14b}) presents the statistics of the root-mean-square (rms) velocity. Compared with the DNS, the SMAG overpredicts the $u_{\text{rms}}$ and underpredicts the $v_{\text{rms}}$ and $w_{\text{rms}}$. The energy from the mean flow is the primary input for the streamwise component $u_{\text{rms}}$, and this energy is redistributed to $v_{\text{rms}}$ and $w_{\text{rms}}$ through pressure. The results of the SMAG indicate that the redistribution mechanism is suppressed. The monoscale and U-Net models slightly overpredict the $u_{\text{rms}}$ and $w_{\text{rms}}$, although not at the scale of the SMAG. By contrast, although the MSC-SGS model slightly overestimates the DNS for the $w_{\text{rms}}$, it predicts the $u_{\text{rms}}$ and $v_{\text{rms}}$ well owing to the appropriate energy redistribution from the $u_{\text{rms}}$ to the $v_{\text{rms}}$ and $w_{\text{rms}}$. Therefore, we can conclude that by considering the interaction between multiscales using the MSC-SGS model, the effect of transferring energy from the large-scale mean flow to the small-scale streamwise vortex is achieved.

\begin{figure*}[ht]
    \centering
    \subfloat[\label{Fig15a}]{\includegraphics[width=0.45\linewidth]{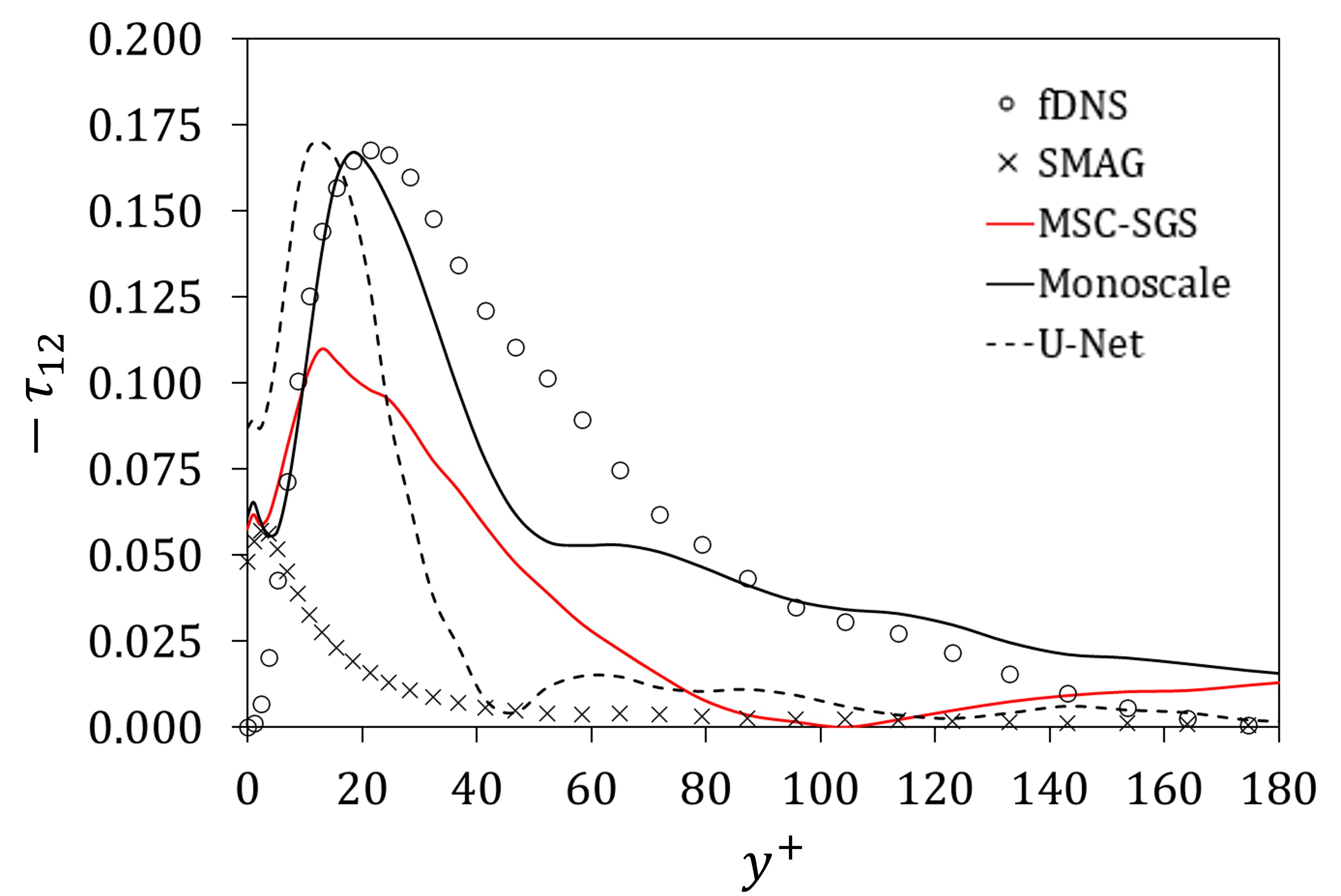}}\hspace{0.5cm}
    \subfloat[\label{Fig15b}]{\includegraphics[width=0.45\linewidth]{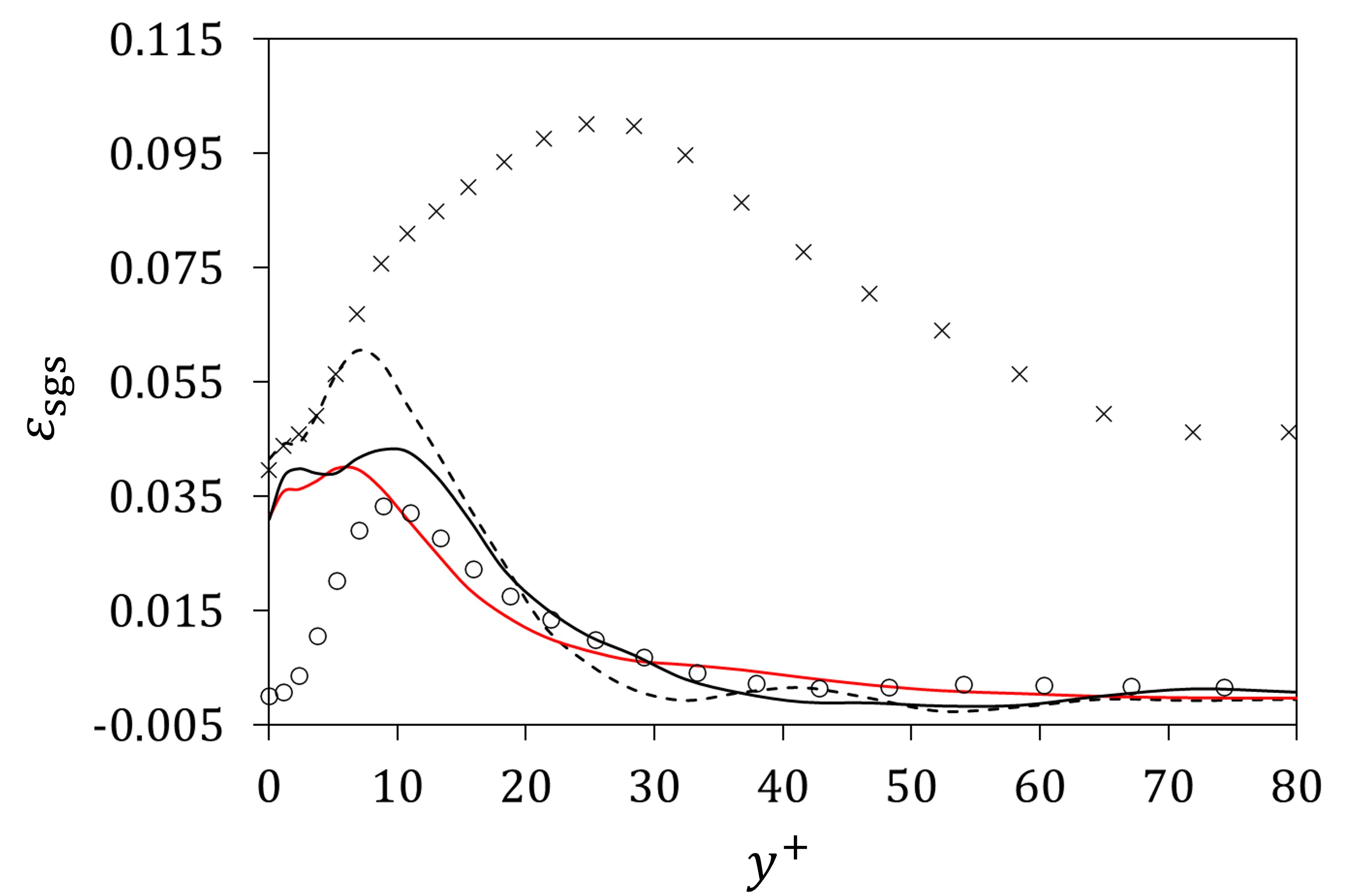}}\\
    \subfloat[\label{Fig15c}]{\includegraphics[width=0.45\linewidth]{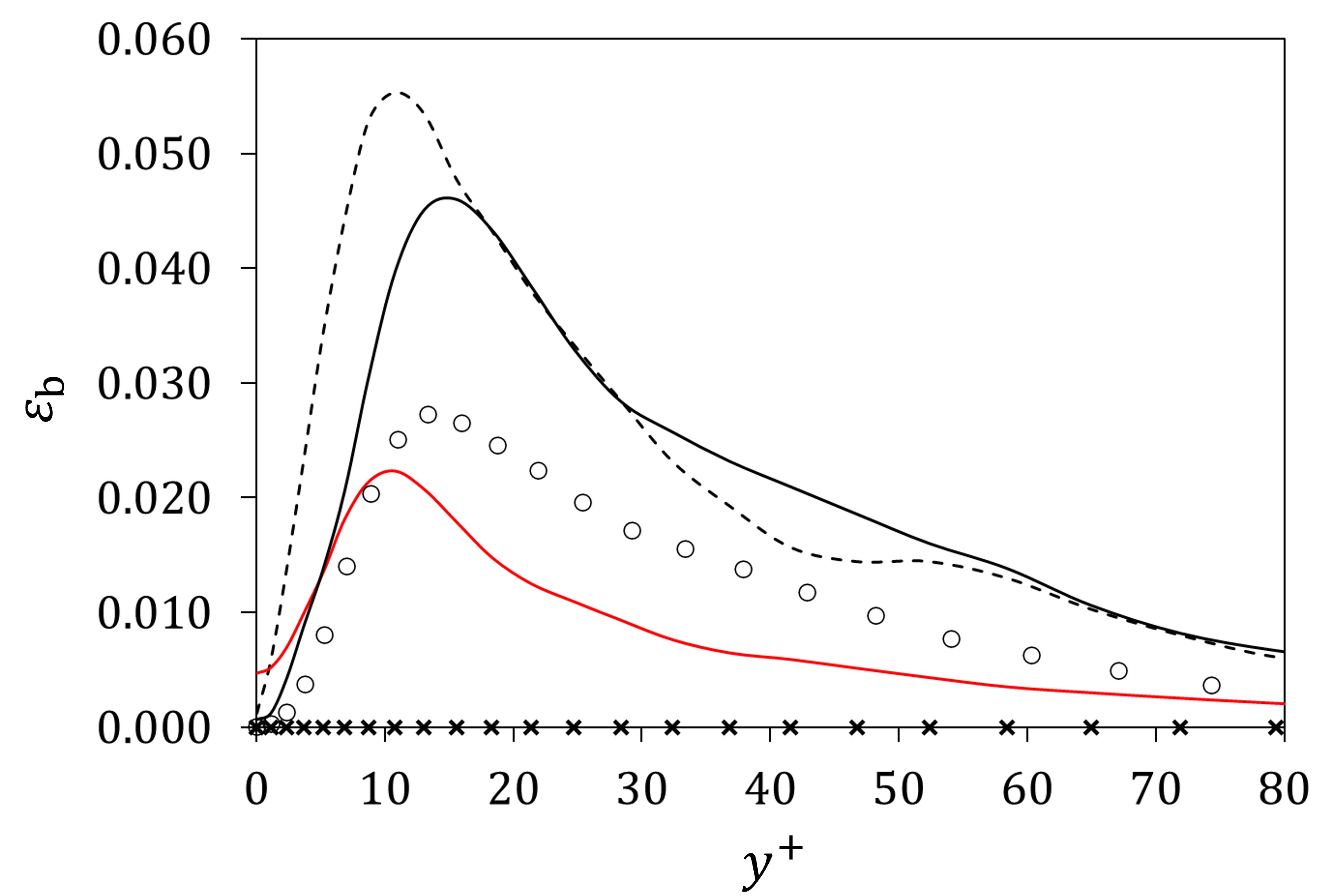}}\hspace{0.5cm}
    \subfloat[\label{Fig15d}]{\includegraphics[width=0.45\linewidth]{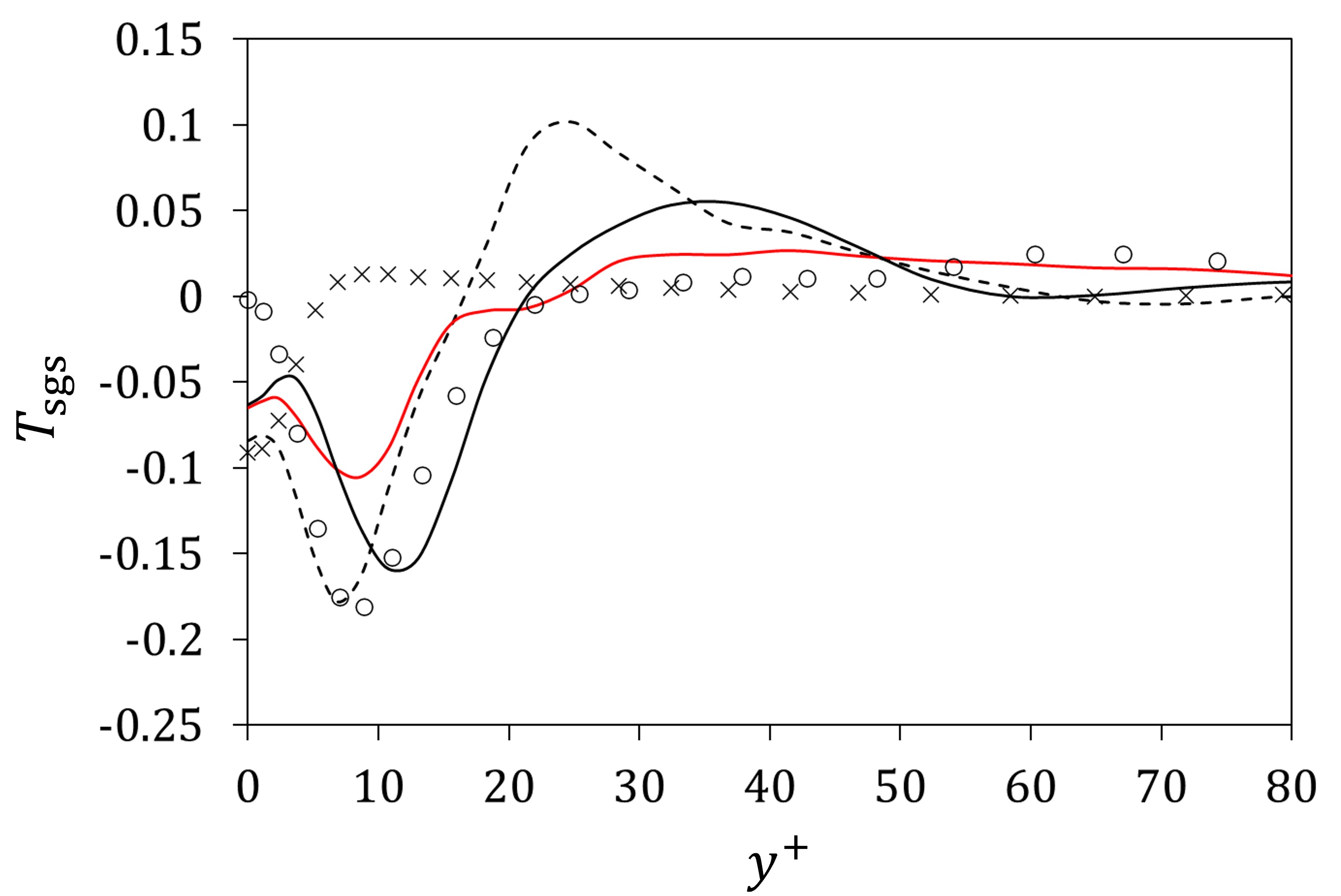}}

    \caption{\textit{A posteriori} test result of wall-normal distribution for (a) $\tau_{12}$, (b) SGS dissipation, (c) SGS backscatter, and (d) SGS transport.}
    \label{Fig15}
\end{figure*}

As shown in Fig. \ref{Fig14}(\subref{Fig14c}), the Reynolds stress deviates significantly from the DNS results. In the near-wall region, the data-driven SGS models align well with the DNS data, whereas the SMAG underpredicts the Reynolds stress. However, these results begin to deviate farther from the wall, thus indicating challenges in accurately capturing the antisymmetric distribution of the Reynolds stress. As mentioned by Kang et al.$^{16}$ and Park \& Choi$^{8}$, this discrepancy is associated with inaccuracies in the SGS shear stress $\tau_{12}$, as highlighted in Fig. \ref{Fig15}(\subref{Fig15a}) which accounts for the inverse relationship between Reynolds stress and $\tau_{12}$. The MSC-SGS model underpredicts $\tau_{12}$ but overpredicts the Reynolds stress, while conversely, the monoscale and U-Net models considerably overestimate $\tau_{12}$, thus resulting in an underestimated Reynolds stress. Meanwhile, for the SMAG, both the Reynolds stress and $\tau_{12}$ are lower compared with the other results. The total shear stress can be expressed as $1 - y/\delta = \text{d}\langle u^+ \rangle/\text{d}y - \langle u' v' \rangle/u_{\tau}^2 - \langle \tau_{12} \rangle/u_{\tau}^2$. When the Reynolds stress and $\tau_{12}$ decrease, the viscous force represented by $\text{d}\langle u^+ \rangle/\text{d}y$ increases to maintain the total shear-stress balance, thus affecting the rate-of-strain tensor $\bar{D}_{ij}$. Fig. \ref{Fig15}(\subref{Fig15b}) shows the dissipation, whereby the SMAG exhibits excessive dissipation, which is consistent with its reliance on a higher velocity gradient to compensate for the total shear stress. These results are consistent with the characteristics of the SMAG reported by Gamahara \& Hattori\cite{gamahara_searching_2017} and Wang et al.\cite{wang_investigations_2018} Fig.\ref{Fig15}(\subref{Fig15b}) shows that both the monoscale and U-Net models exhibit higher dissipation than the fDNS. By contrast, the MSC-SGS model aligns better with the DNS data across most regions, which is critical for accurately capturing the energy transfer between scales, except near the wall. In the near-wall region, all SGS models show excessive dissipation, which may compromise the near-wall viscous effects and contribute to the slight discrepancies observed in the mean velocity profile (Fig. \ref{Fig14}(\subref{Fig14a})).

\begin{figure}
\includegraphics[width=1\linewidth]{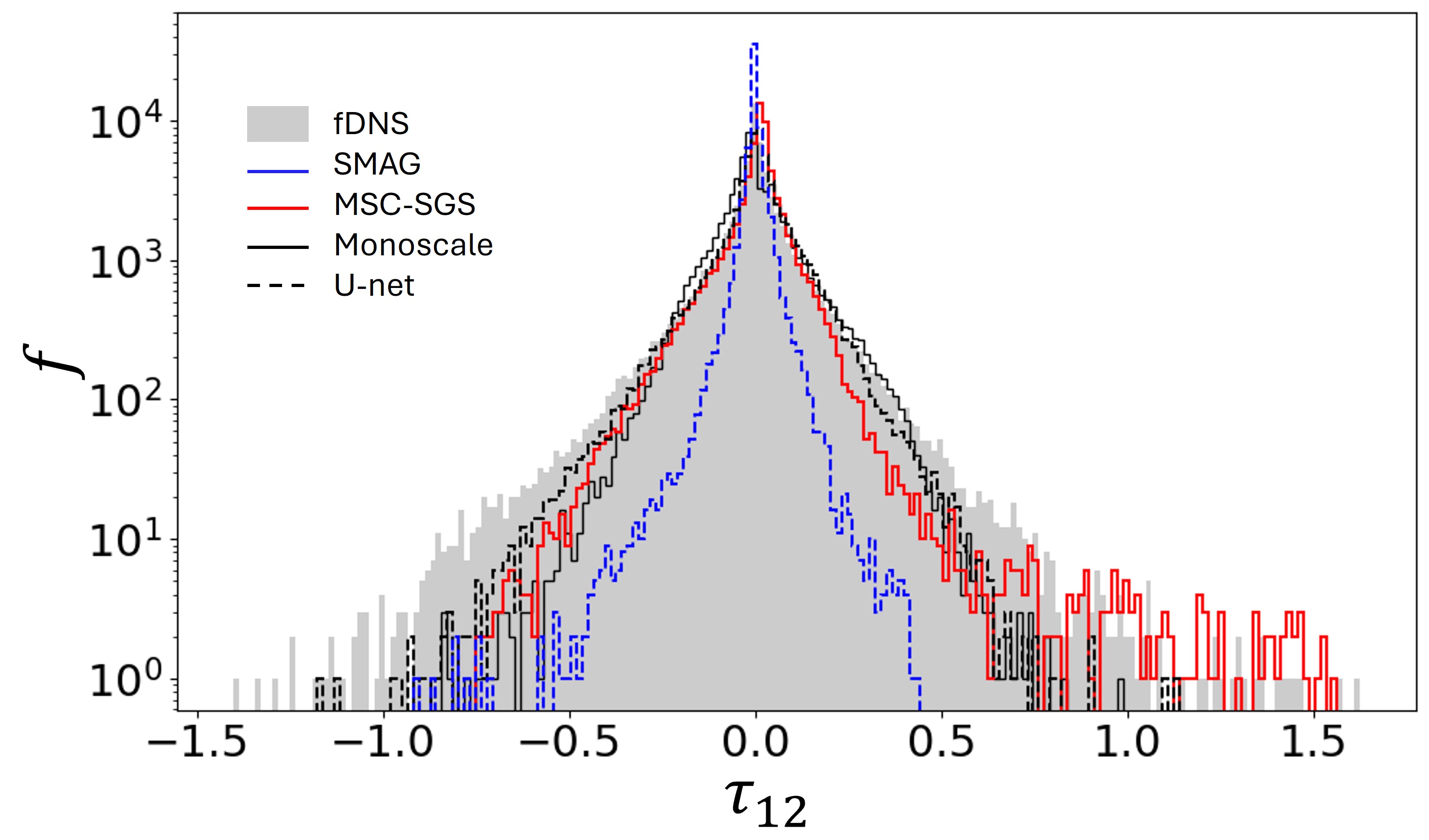}
\caption{\label{Fig16} {Distribution of number of occurrences ($f$) of $\tau_{ij}^{fDNS}$ and $\tau_{ij}^{p}$ for $\tau_{12}$.}}
\end{figure}

\begin{figure*}[ht]
    \centering
    \subfloat[]{%
        \includegraphics[width=0.4\linewidth]{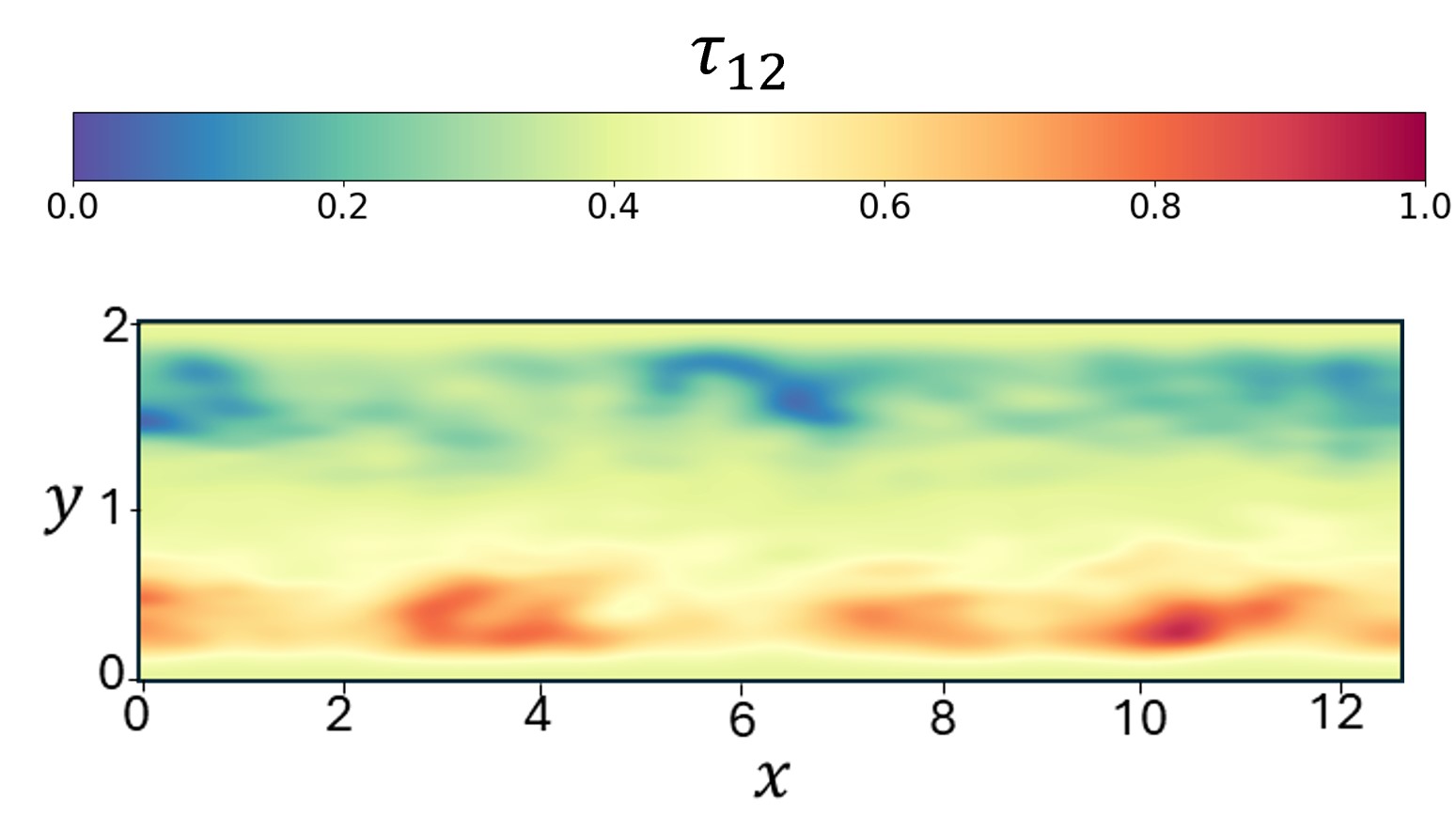}} \hspace{0.5cm} \\
    
    \subfloat[]{%
        \includegraphics[width=0.4\linewidth]{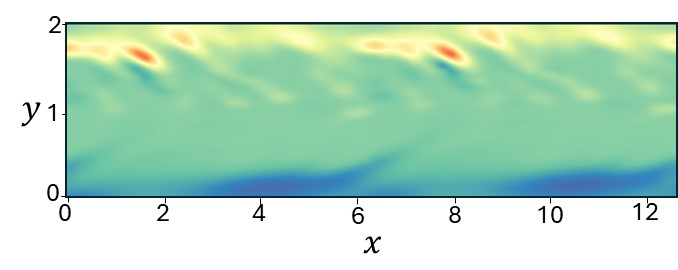}} \hspace{0.5cm}
    \subfloat[]{%
        \includegraphics[width=0.4\linewidth]{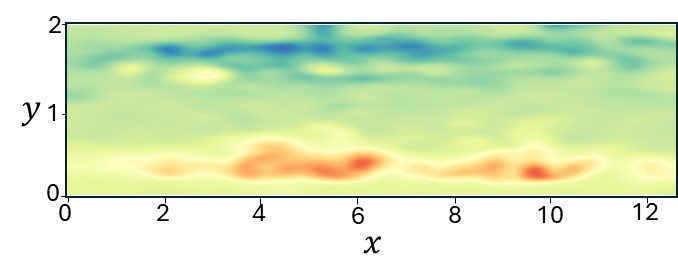}} \\

    \subfloat[]{%
        \includegraphics[width=0.4\linewidth]{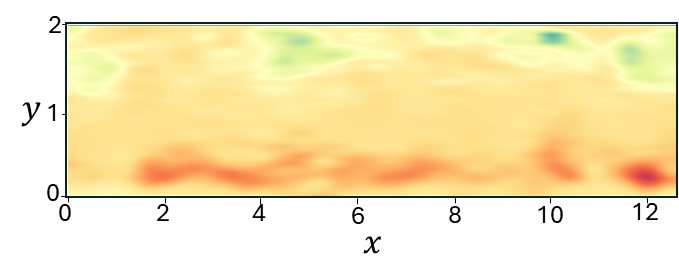}} \hspace{0.5cm}
    \subfloat[]{%
        \includegraphics[width=0.4\linewidth]{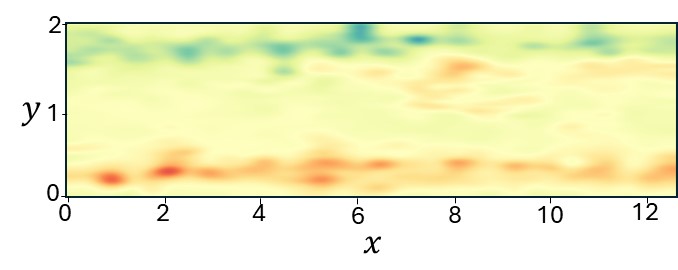}} 

    \caption{Instantaneous $\tau_{12}$ in $x-y$ plane at $z=L_{z}/2$ for (a) fDNS, (b) SMAG, (c) MSC-SGS, (d) Monoscale, and (e) U-Net. Value is normalized to be within $[0, 1]$.}
    \label{Fig17}
\end{figure*}

\begin{figure}
\includegraphics[width=1\linewidth]{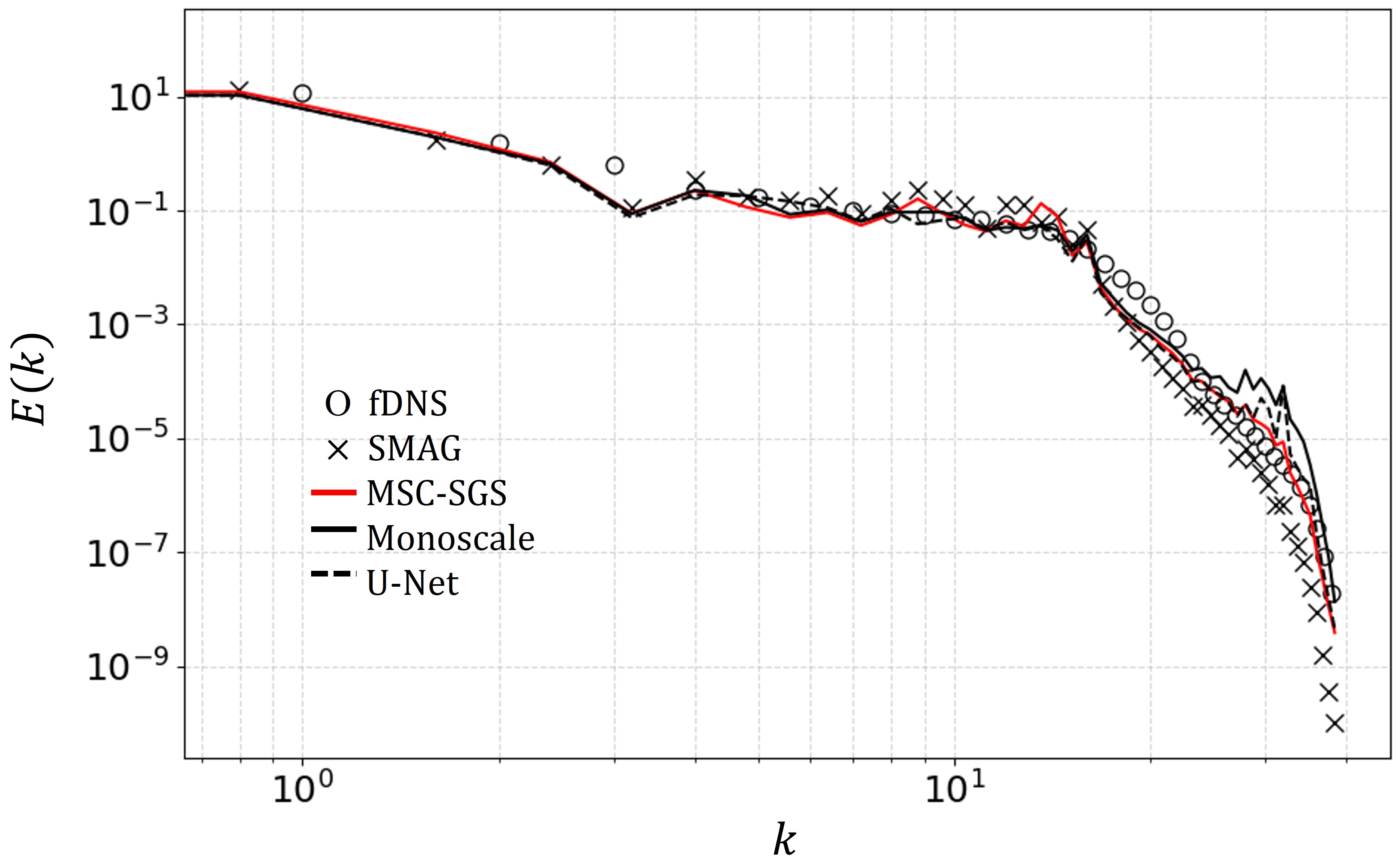}
\caption{\label{Fig18} {Energy spectra of velocity fluctuations.}}
\end{figure}

\begin{figure*}[ht]
    \centering
    \includegraphics[width=0.4\linewidth]{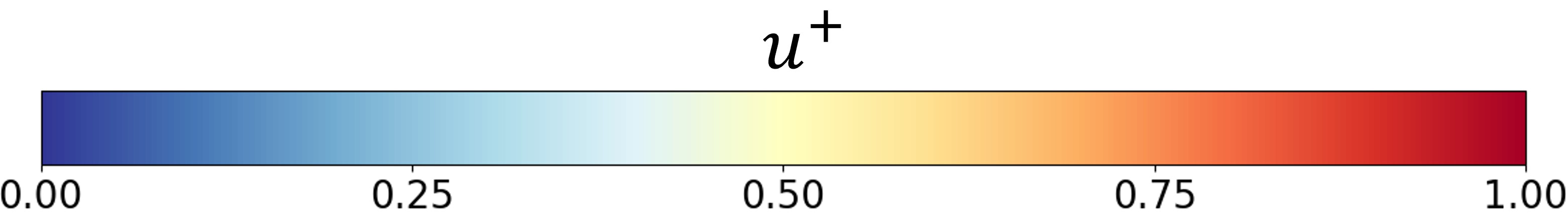} \\
        
    \subfloat[]{%
        \includegraphics[width=0.4\linewidth]{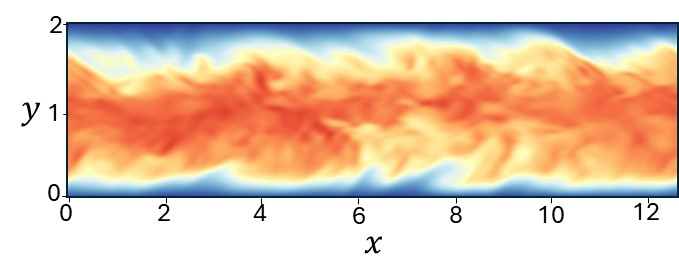}} \hspace{0.5cm}
    \subfloat[]{%
        \includegraphics[width=0.4\linewidth]{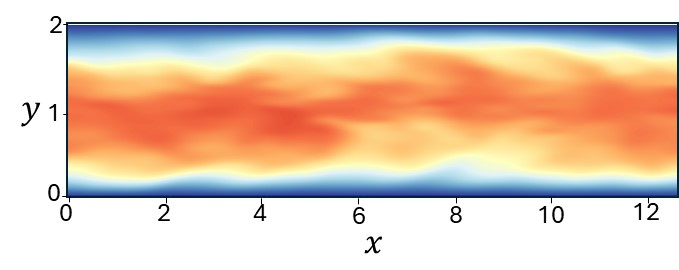}} \\

    \subfloat[]{%
        \includegraphics[width=0.4\linewidth]{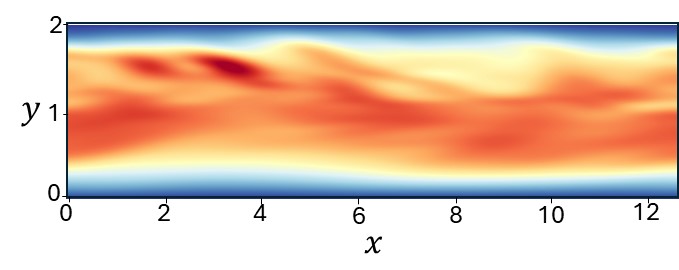}} \hspace{0.5cm}
    \subfloat[]{%
        \includegraphics[width=0.4\linewidth]{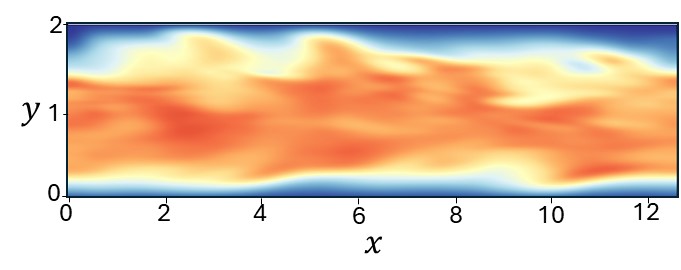}} \\

    \subfloat[]{%
        \includegraphics[width=0.4\linewidth]{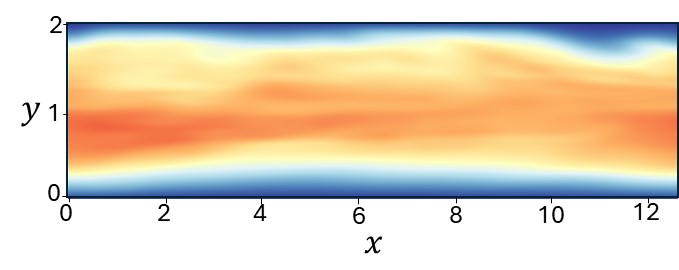}} \hspace{0.5cm}
    \subfloat[]{%
        \includegraphics[width=0.4\linewidth]{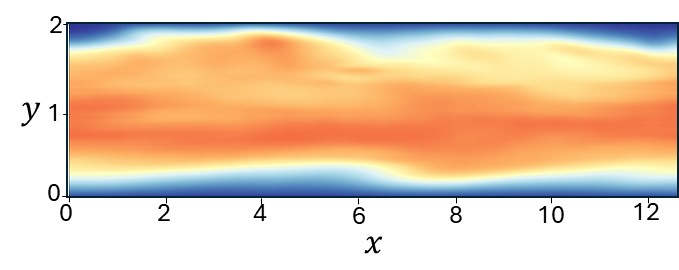}}

    \caption{Comparison of streamwise velocity between DNS and LES results at $z = L_z/2$. (a) DNS, (b) fDNS, (c) SMAG, (d) MSC-SGS, (e) Monoscale, and (f) U-Net.}
    \label{Fig19}
\end{figure*}

\begin{figure*}[ht]
    \centering
  
    \subfloat[]{%
        \includegraphics[width=0.4\linewidth]{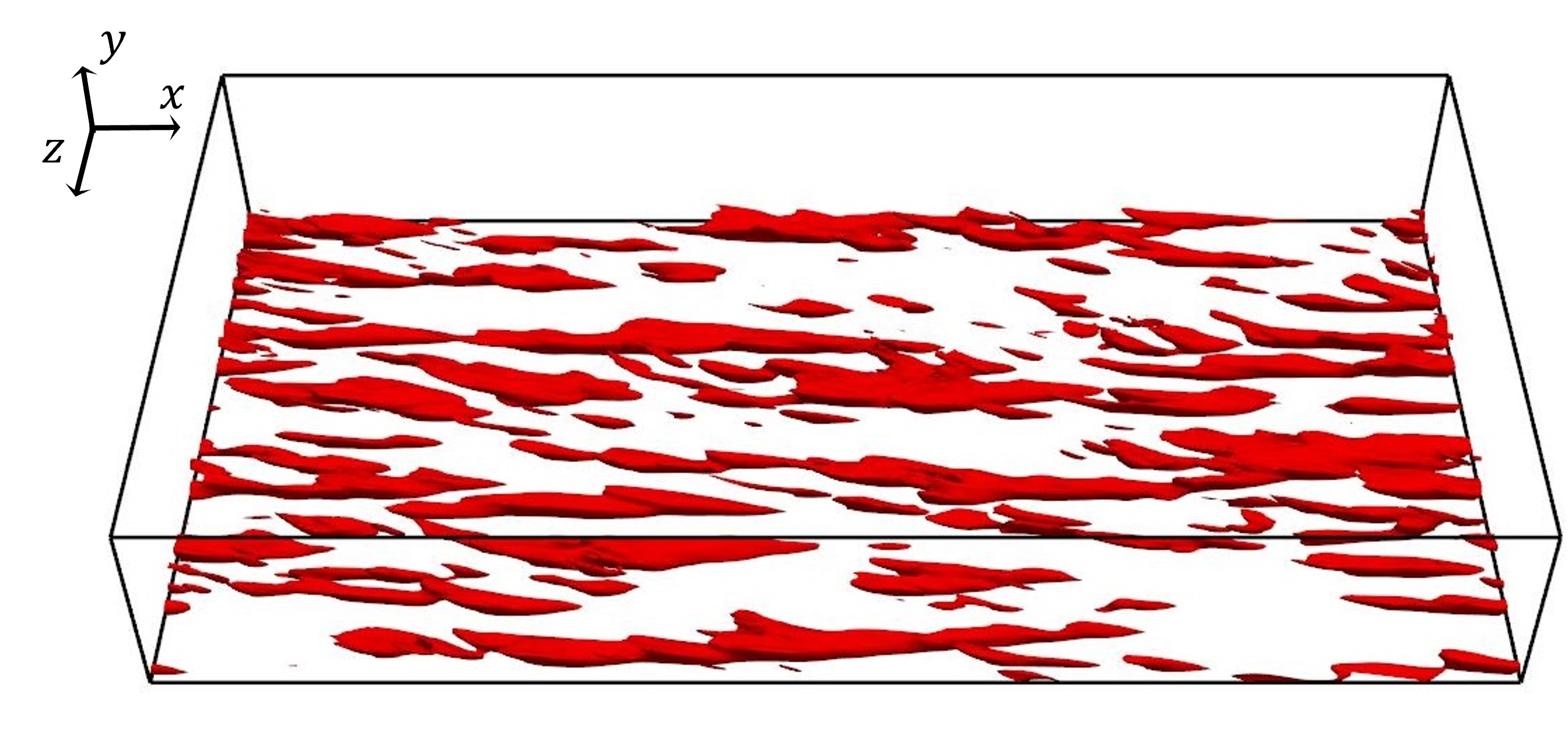}} \hspace{0.5cm}
    \subfloat[]{%
        \includegraphics[width=0.4\linewidth]{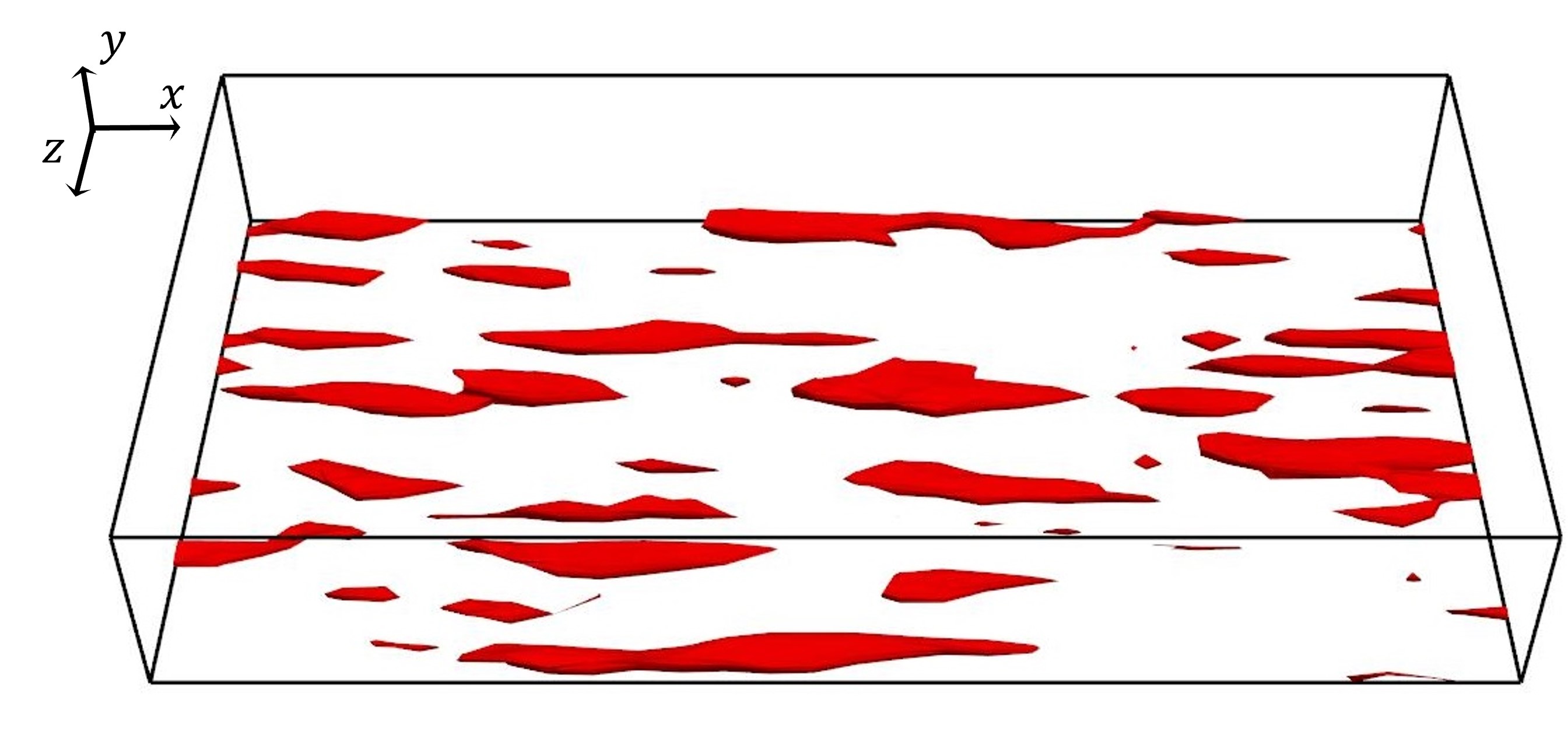}} \\

    \subfloat[]{%
        \includegraphics[width=0.4\linewidth]{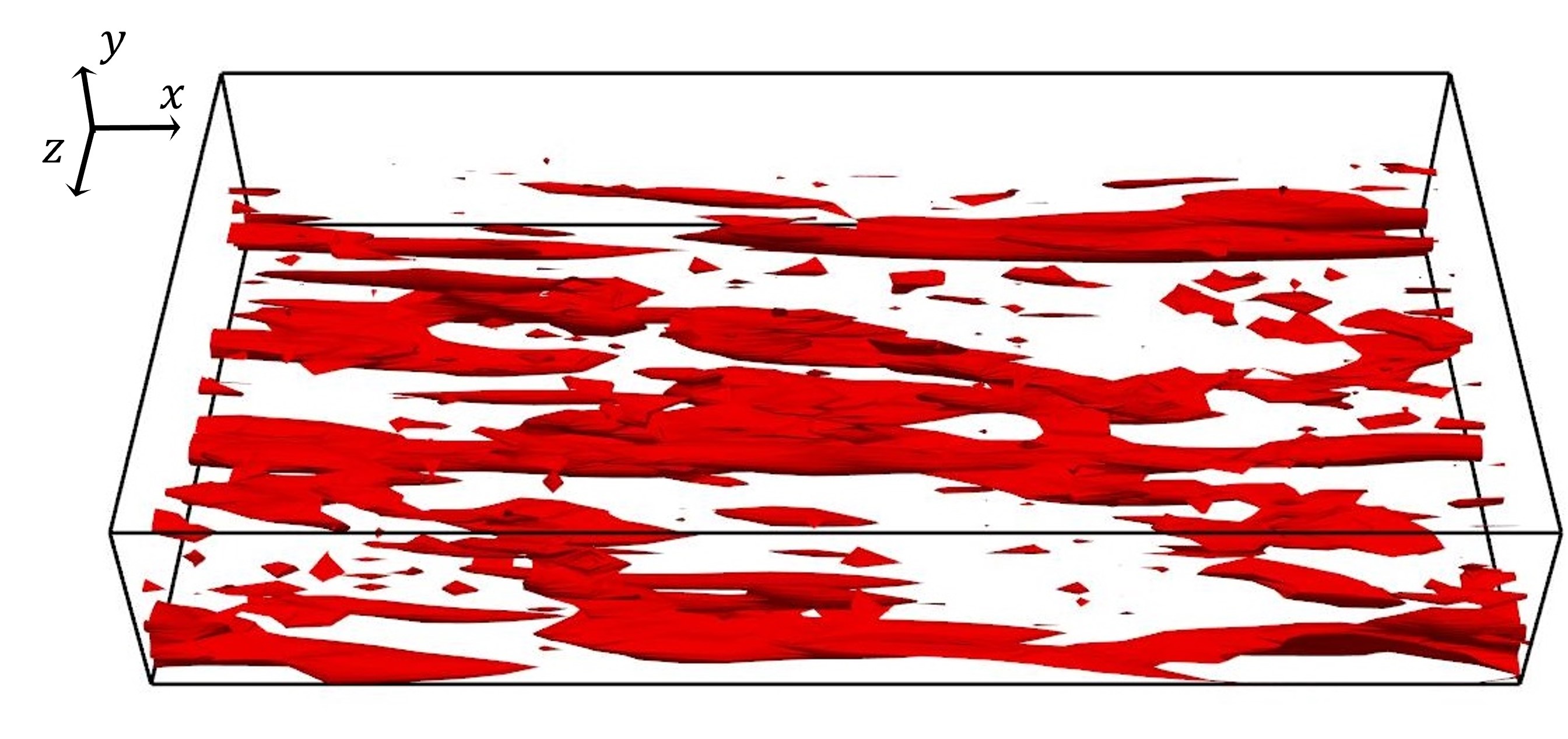}} \hspace{0.5cm}
    \subfloat[]{%
        \includegraphics[width=0.4\linewidth]{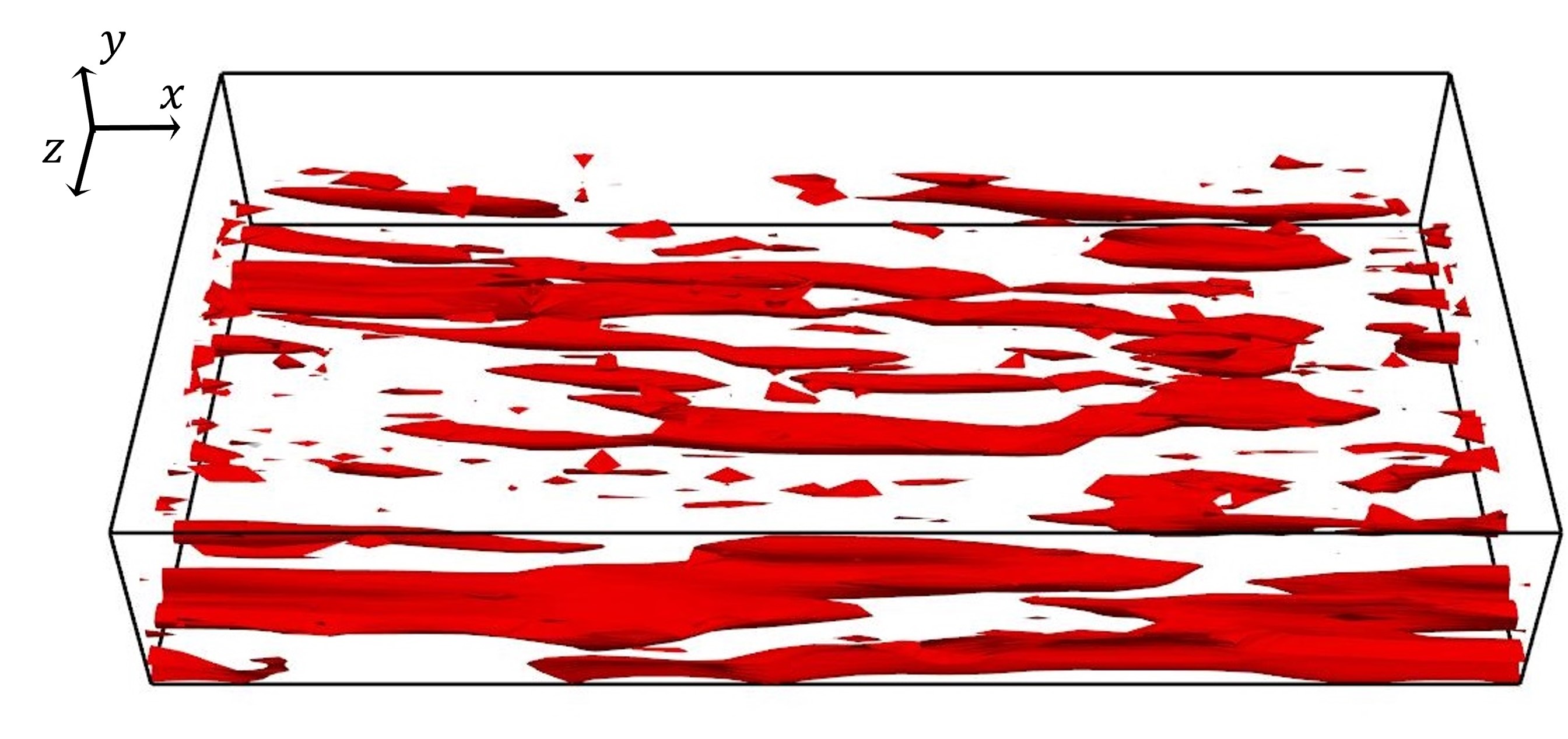}} \\

    \subfloat[]{%
        \includegraphics[width=0.4\linewidth]{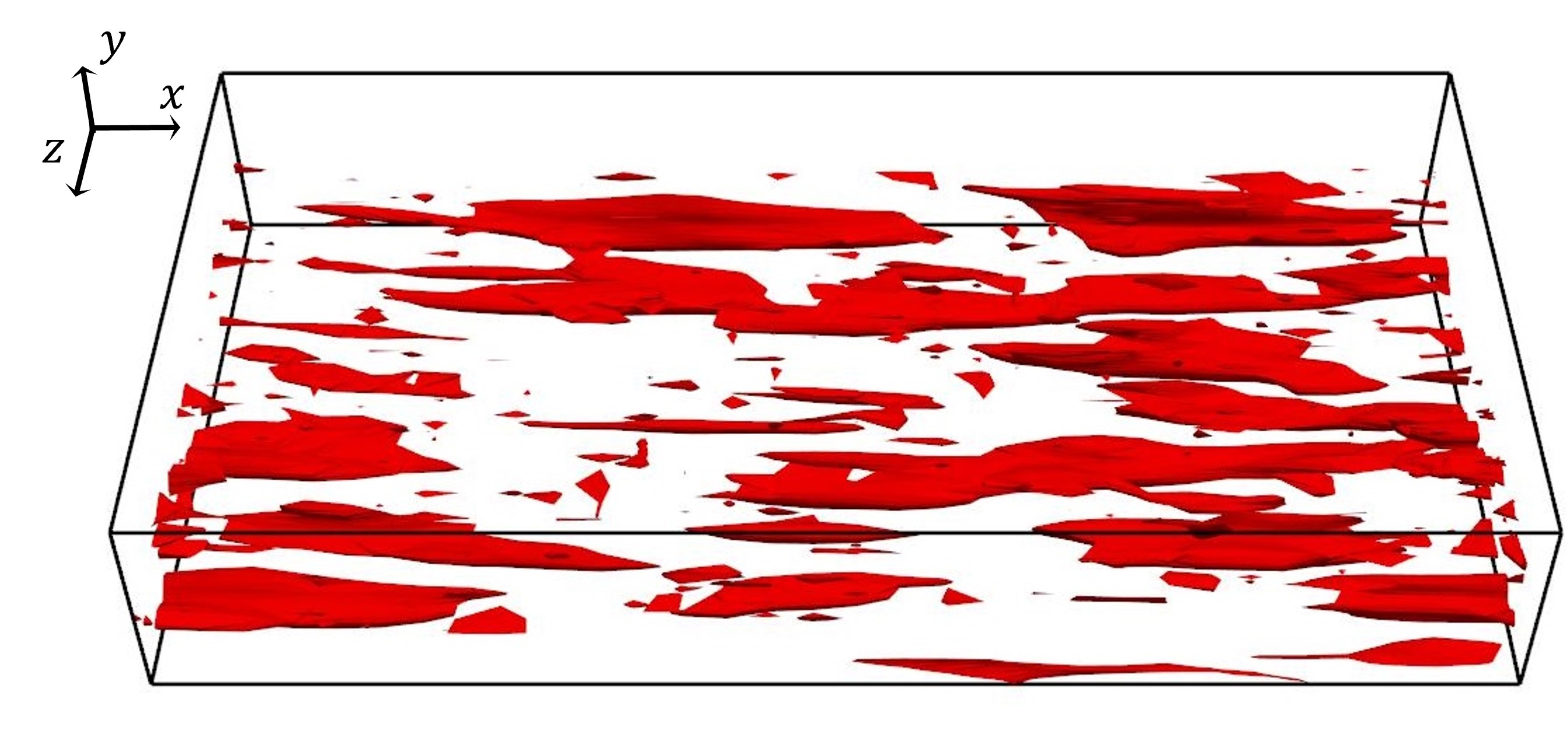}} \hspace{0.5cm}
    \subfloat[]{%
        \includegraphics[width=0.4\linewidth]{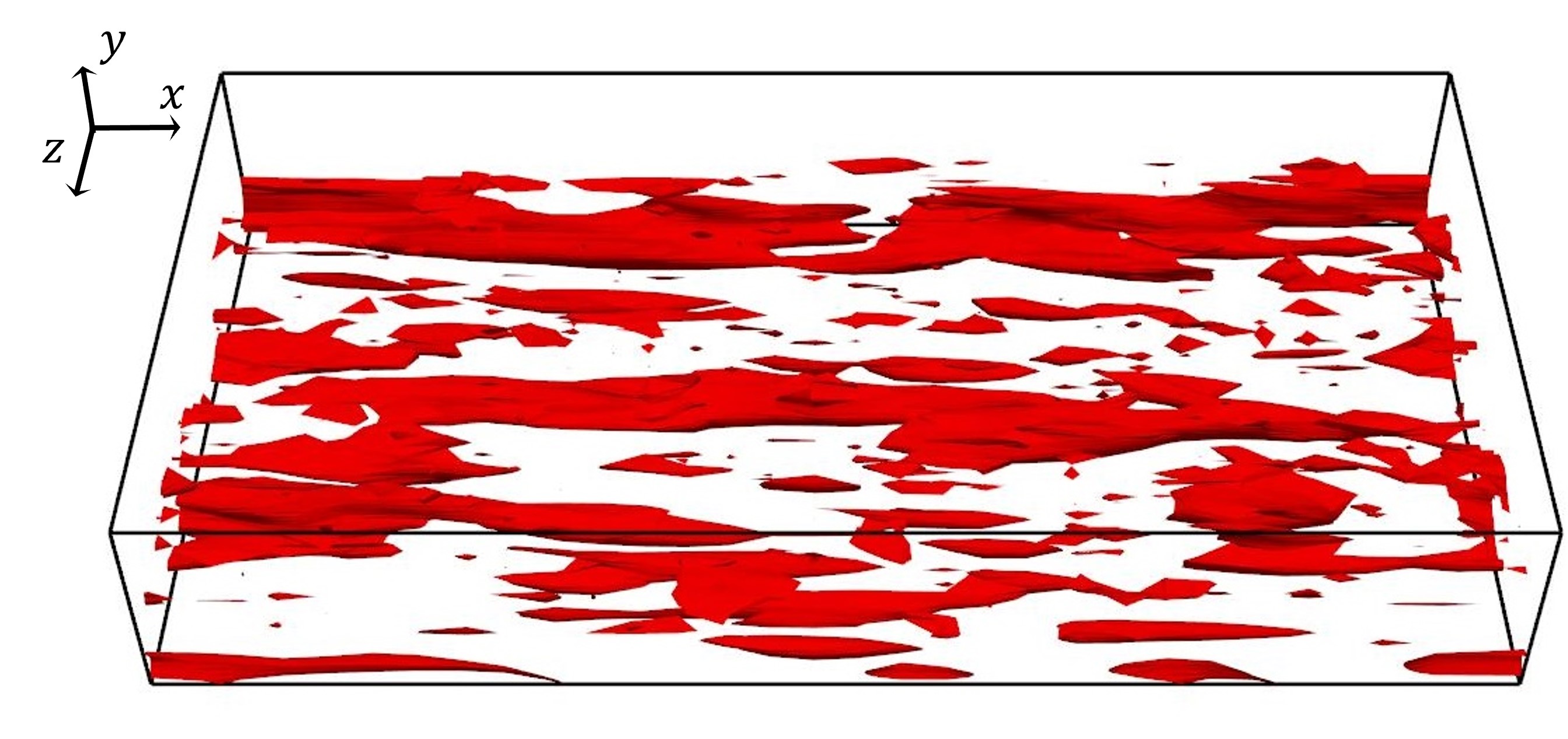}}

    \caption{Comparison of isosurfaces of $Q^+ = 0.2 / Re_{\tau}^2$ between DNS and LES model. (a) DNS, (b) fDNS, (c) SMAG, (d) MSC-SGS, (e) Monoscale, and (f) U-Net.}
    \label{Fig20}
\end{figure*}

Figure \ref{Fig15}(\subref{Fig15c}) shows the backscatter in the \textit{a posteriori} test. Excessive backscatter amplifies the existing velocity fluctuations, increases the energy at resolved scales, and may destabilize the numerical simulation. Guan et al.$^{9}$ described the advantages of CNN-based models, which can effectively capture backscatters, while maintaining numerical stability. Here, the monoscale and U-Net models show considerably overpredicted backscatter, which results in numerical inaccuracy, as shown in Fig. \ref{Fig14}. The MSC-SGS model may underpredict the backscatter to some extent; however, its predictions remain relatively close to those of the fDNS, thus resulting in improved numerical accuracy compared with the other SGS models. These findings are consistent with those of Park \& Choi\cite{park_toward_2021} and Kang et al.\cite{kang_neural-network-based_2023}, which emphasized that models capable of capturing inverse cascades without severe overestimation ensured both numerical stability and accuracy. Conversely, the SMAG cannot represent the backscatter because of its local equilibrium assumption. The capability of the MSC-SGS model to represent backscatter while maintaining stability and accuracy underscores its ability to simulate turbulent flows.

Figure \ref{Fig15}(\subref{Fig15d}) summarizes \textit{a posteriori} results for SGS transport $T_{sgs}$. The SMAG exhibited low transport values, thus indicating minimal energy transfer at the SGS, whereas most of the energy is dissipated. Meanwhile, the monoscale and U-Net models overestimate $T_{sgs}$ in the $20 < y^+ < 40$ region, thus implying excessive energy transfer. These results align with Fig.~15(c), which shows that their contribution to backscatter is exaggerated compared with that of the fDNS. The MSC-SGS model demonstrates a more accurate prediction of $T_{sgs}$ across most regions, except for a deviation near the wall. Its transport values are consistent with the fDNS data, thus reflecting a balanced energy redistribution. In general, the ability of the MSC-SGS model to moderately represent backscatter and capture dissipation ensures that the energy-transfer processes are well modeled. These results are consistent with those of the \textit{a priori} test, thus underlining the importance of accurately modeling $T_{sgs}$ to achieve a reliable LES, as reported by Volker et al.\cite{volker_optimal_2002}


Figure \ref{Fig16} presents a comparison of the number of occurrences of $\tau_{12}$ data for the SGS model. Here, the number of bins in the histogram is set to 200, which corresponds to the bin width of $0.01$. The $\tau_{12}$ data within each bin is counted. Based on the histogram, the MSC-SGS model has a better distribution than the monoscale and U-Net models. It aligns closely with the fDNS, particularly at the peak and positive tails of the distribution, while underpredicting the negative value. Meanwhile, the SMAG model presents a notable limitation, as it predominantly represents lower values of $\tau_{ij}$, which agrees well with Fig. \ref{Fig15}(\subref{Fig15a}). The discrepancy in the negative tails is attributable to the limited ability of the LES model in accurately representing the momentum transfer near the wall. In addition, the qualitative result of instantaneous $\tau_{12}$ is presented in Fig. \ref{Fig17}. The SMAG model offers a lower distribution as it underestimates $\tau_{12}$. The MSC-SGS model shows a more detailed representation of $\tau_{12}$ with higher distribution values, as it tends to yield more positive values, as shown in Fig. \ref{Fig16}. The MSC-SGS model provides a more reasonable representation compared with the fDNS. Both the monoscale and U-Net models also show good representations, as shown in Fig. \ref{Fig17}, which align well with the trends shown in Figs. \ref{Fig15}(\subref{Fig15a}) and \ref{Fig16}. In general, all DNN-based models reproduce the $\tau_{12}$ field better than the conventional SMAG, thus highlighting their potential for accurately capturing SGS stresses.

Figure \ref{Fig18} shows the 3D energy spectra $E(k)$ of the velocity fluctuations. In the low-wavenumber region, which corresponds to a large energy-containing scale, all models agree well with the fDNS data. The dissipative scale of the small-scale turbulence is represented by the high-wavenumber region. Here, the SMAG model is underperformed by exhibiting lower energy values compared with the fDNS results. This indicates that excessive dissipation results in kinetic energy loss, thus further confirming the limitation of the SMAG model in representing the dynamics at smaller scales (Fig. \ref{Fig15}(\subref{Fig15b})). Meanwhile, the monoscale and U-Net models exhibit a slight energy pile-up near the dissipation range. This excess energy at high wavenumbers suggests difficulties in accurately resolving small-scale turbulence. Furthermore, these models present limitations in capturing the energy-transfer process, particularly in the dissipative range. Compared with other $\tau_{ij}^{DNN}$ model, the MSC-SGS model aligns well with the fDNS data in all wavenumber regions. By maintaining a closer match with the fDNS data, it underscores the importance of multiscale representation for capturing a wide range of turbulent scales.

The contours of the instantaneous streamwise velocity component ($x$-$y$ plane) are shown in Fig. \ref{Fig19}. Because of the filtering operation, small-scale vortices disappear in the fDNS, thus causing only larger vortices to remain. In general, the LES models capture the overall distribution of the streamwise velocity. The MSC-SGS model shows a velocity distribution that aligns closely with the DNS and fDNS results, where the overall flow characteristics are effectively captured and matched the mean velocity, as shown in Fig. \ref{Fig14}(\subref{Fig14a}). The SMAG model provides a good representation but exhibits a higher velocity in some regions, thereby indicating stronger streamwise velocity fluctuations. Meanwhile, the monoscale and U-Net models show a more homogenized pattern with less detailed flow structures, particularly at the centerline, which represents a lower mean velocity, as shown in Fig. \ref{Fig14}(\subref{Fig14a}). Fig. \ref{Fig20} illustrates the instantaneous vortical structures described by the Q-criterion ($Q=-\frac{1}{2} \frac{\partial u_i}{\partial x_j} \frac{\partial u_j}{\partial x_i}$). The fDNS data shows a relatively large scale, whereas the DNS represents all scales. In the LES, both the SMAG and $\tau_{ij}^{DNN}$ model depict large, elongated tube-like coherent vortices. Visually, the SMAG model shows the dominance of elongated coherent vortices. For the $\tau_{ij}^{DNN}$ model, all models demonstrate the ability to capture small and large scales. Compared with the results of Park \& Choi$^8$ and Kang et al.$^{16}$, which exhibited fewer large vortices, the current models present the dominance of large vortices. This may reflect the limitations of the model in accurately resolving small-scale vortices near the wall, which warrants further investigation.

\section{Conclusion}
In this study, we investigated an SGS model for an LES using a data-driven SGS approach. The multiscale algorithm of the MSC-SGS model was employed to predict the SGS stresses $\tau_{ij}^{p}$. This model encoded multiscale input representations obtained through low-pass filtering at different scales, thus allowing it to effectively capture interactions between scales. By progressively encoding features from large to small scales, it captured the energy-transfer process, which was analogous to the energy cascade, thus ensuring comprehensive detail retention for the accurate prediction of residual stress. This approach enhanced the model’s ability to capture both large- and small-scale features, thereby demonstrating its proficiency in nonlinear DNN regression models.

\textit{A priori} results showed that the MSC-SGS model consistently outperformed the monoscale and U-Net algorithms by achieving a higher CC across all wall regions. The MSC-SGS model improved the prediction accuracy, as indicated by the average CC value, with significant increases by 7.8\% and 9.4\% for smaller stress components $\tau_{13}$  and $\tau_{23}$, respectively. Physical quantity analysis showed that all models exhibited good predictions for shear stress, SGS dissipation, SGS backscatter, and SGS transport. In terms of robustness, all models maintained a stable CC value over the temporal evolution. Although the MSC-SGS model exhibited slight sensitivity to noise, it consistently outperformed the other models under dynamic and noisy conditions.

Meanwhile, \textit{a posteriori} results showed that the LES computation based on the MSC-SGS model agreed well with the DNS data in terms of turbulence statistics. Additionally, for the Reynolds stress prediction, the MSC-SGS model showed alignment near the wall but overpredicted far from the wall, whereas the SMAG, monoscale model, and U-Net model consistently underpredicted in the $y^{+}>30$ region. A multiscale analysis of the energy spectra revealed the ability of various SGS models to capture turbulence structures across a wide range of scales. All models aligned well with the fDNS data at large scales, thus demonstrating their capability to accurately resolve large-scale turbulent structures. However, the MSC-SGS model maintained close agreement with the fDNS data across all wavenumber regions, thereby underscoring its superiority in multiscale representation and its ability to capture the full spectrum of turbulent dynamics. By contrast, the SMAG underperformed at high wavenumbers, where it exhibited lower energy values owing to kinetic-energy loss attributed to excessive dissipation. The monoscale and U-Net models showed an energy pile-up near the dissipation range, thus suggesting difficulties in accurately resolving small-scale turbulence and energy-transfer processes. 

Despite its advantages, the MSC-SGS model presented some limitations, particularly in the near-wall region. Future studies shall focus on improving the model’s ability to address near-wall problems. More importantly, the generalizability should be further investigated to ensure the applicability of the model to diverse flow conditions beyond the training datasets. Additionally, the computational cost of the MSC-SGS model remains a significant concern because the multiscale decomposition and DNN inference require longer simulation times compared with the case of conventional models such as the SMAG. Future studies should aim to optimize the model architecture to reduce the computational overhead without sacrificing accuracy.

\begin{acknowledgments}
This work was partly supported by the Research Proposal-based Use of the Project for Nurturing Student Competing with the World at the Large-Scale Computer System-D3 Center, Osaka University. This study was financially supported by JSPS KAKENHI grant No. JP22K03925 and a grant from Beasiswa Pendidikan Indonesia (BPI)/ LPDP (the Indonesian Endowment Fund for Education, Ministry of Finance of the Republic of Indonesia).
\end{acknowledgments}

\section{REFERENCES}
\bibliographystyle{aipnum4-1}
\bibliography{aipsamp}

\end{document}